\documentclass[letterpaper,twocolumn,10pt]{article}
\usepackage{usenix,endnotes,color,pseudocode}

\usepackage{ifpdf}
\ifpdf
\else
\fi

\usepackage{tabularx}
\usepackage{graphicx}
\usepackage{subfigure}
\usepackage[ruled,vlined]{algorithm2e}
\usepackage{enumitem}
\usepackage{authblk}
\usepackage{breakurl}
\usepackage[hyphens]{url}
\setlist{nolistsep,leftmargin=*}
\usepackage{balance}

\usepackage{breakurl}
\usepackage{color}
\usepackage{comment}
\usepackage{multirow}

\usepackage{amssymb,amsmath,amsthm}

\numberwithin{subcase}{case}

\newcommand{\http}{HTTP}
\graphicspath{{figure/}}

\newcommand{\rate}{\ensuremath{\mathrm{Rate}}}

\newcommand{\buf}{\ensuremath{\mathrm{Buf}}}
\newcommand{\prev}{\ensuremath{\mathrm{prev}}}
\newcommand{\nxt}{\ensuremath{\mathrm{next}}}
\newcommand{\now}{\ensuremath{\mathrm{now}}}

\usepackage{leading}
\leading{12.08pt}

\begin{document}
\date{}

\title{Using the Buffer to Avoid Rebuffers:\\ Evidence from a Large Video Streaming Service}

\author[1, 2]{\rm Te-Yuan Huang}
\author[1]{\rm Ramesh Johari}
\author[1]{\rm Nick McKeown}
\author[2]{\rm Matthew Trunnell}
\author[2]{\rm Mark Watson}
\affil[1]{Stanford University}
\affil[2]{Netflix}



\maketitle

\begin{abstract}
To provide a better streaming experience, video clients today select their video rates by observing and estimating the available capacity. 
Recent work has shown that capacity estimation is fraught with difficulties because of 
complex interactions between the ABR control loop, HTTP server performance and TCP congestion control.
Estimation-based rate selection algorithms can lead to unnecessary rebuffering events and suboptimal video quality. 
This paper argues that we should do away with estimating network capacity, and instead directly observe and control the playback buffer---which is the state variable we are most interested in controlling. We present a class of ``buffer-based'' rate selection algorithms that reduce the rebuffering rate while allowing us to control the delivered video quality. 
We implemented our algorithms inside the Netflix video client and ran a series of experiments spanning millions of Netflix users around the world. 
Our results show that by doing away with estimating network capacity and instead focusing on buffer occupancy, we can reduce rebuffer rates by 20\% while holding video rate constant.  
\end{abstract}



\section{Introduction}
\label{sec:intro}


During peak viewing time, well over 50\% of US Internet traffic is streamed
video from Netflix and YouTube~\cite{sandvine-report}. Each video is encoded at
a number of different rates (typically 235kb/s SD to 5Mb/s HD) and stored on
servers.  The video client -- running on a home TV, game console, web browser,
DVD player, etc. -- chooses which video rate to stream by monitoring network
conditions and estimating the available network capacity. The process is usually
called {\em adaptive bit rate selection} or ABR.


\begin{figure}
\centering
\includegraphics[width=0.5\linewidth]{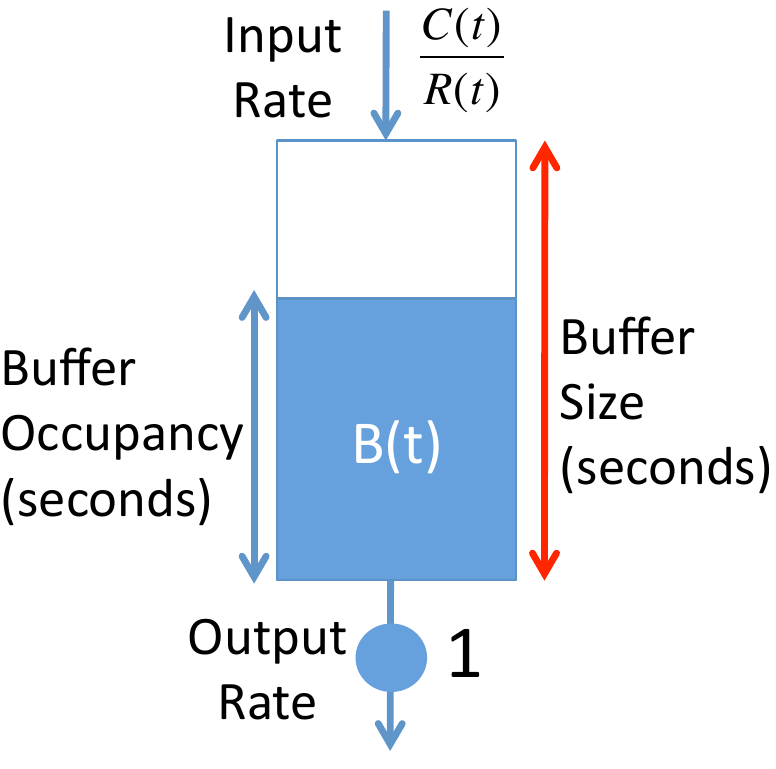}
\caption{The relationship between system capacity, C(t),  and video rate, R(t), in a video playback buffer.}
\label{fig:buffer_ideal_model}
\end{figure}

In this paper, we study the performance of ABR in the Netflix video streaming service.  ABR algorithms used by such services balance two overarching goals.  On one hand, they try to maximize the video quality by picking the highest
video rate the network can support.  Second, they try to minimize {\em rebuffering events} which cause the video to halt if the client's playback buffer goes empty. 

It is easy for a streaming service to meet either one of the objectives on its own.  To maximize video quality, a service could just stream at the maximum video rate $R_{\max}$ all the time.  Of course, this would risk extensive rebuffering.  On the other hand, to minimize rebuffering, the service could just stream at the minimum video rate $R_{\min}$ all the time.  Of course, this would lead to relatively low video quality.  The design goal of a video streaming service is to {\em simultaneously} obtain high performance on both metrics.

Our key contribution is to propose a new class of ABR algorithms that pick a video rate based solely on the occupancy of the playback buffer. Our approach has the intuitively appealing property of directly tracking the buffer (i.e., the state variable) we are trying to control. The approach is in stark contrast to existing ABR
algorithms that base their choice on estimates of available network
capacity.

We find that {\em buffer-based ABR algorithms can reduce the rebuffering rate by 20\% compared to existing best-of-breed ABR algorithms, while delivering a similar video rate.}  We make this observation  based on a sequence of live experiments conducted over millions of randomly selected users on three continents during May-September 2013 on the Netflix video streaming service.

Why should we expect it to be possible to improve the rebuffering rate at all?  
To understand rebuffering it helps to study the dynamics of the playback buffer in the client, shown in Figure~\ref{fig:buffer_ideal_model}. We track the buffer occupancy in {\em seconds of video}. Every second, one second of video is removed from the buffer and played to the user. The buffer drains at unit rate (since one second is played back every second of real time). The client requests {\em chunks} of video from the server, each chunk containing a fixed duration of video (four seconds per chunk). The higher the video rate, the larger the chunk (in bytes). 

If the ABR algorithm picks a video rate, $R(t)$, that is greater than the network capacity, $C(t)$, then new data is put into the buffer at rate $C(t)/R(t) < 1$ and so the buffer decreases. Put another way, if more than one chunk is played before the next chunk arrives, then the buffer is depleted. If the ABR algorithm keeps requesting chunks that are too big  for the network to sustain (i.e. the video rate is too high), eventually the buffer will run dry, playback freezes and we see the familiar ``Rebuffering...'' message on the screen.  
 
Notice that if the network capacity is always greater than the lowest video rate $R_{\min}$, i.e. $C(t) > R_{\min}, \forall t>0$,  there {\em never} needs to be a rebuffering event --- the algorithm can simply pick $R(t) = R_{\min}$ so that $C(t)/R(t) > 1, \forall t>0$ and the buffer keeps growing. In this case, rebuffering is unnecessary and should never happen.

\begin{figure}
\centering
\includegraphics[width=0.9\linewidth]{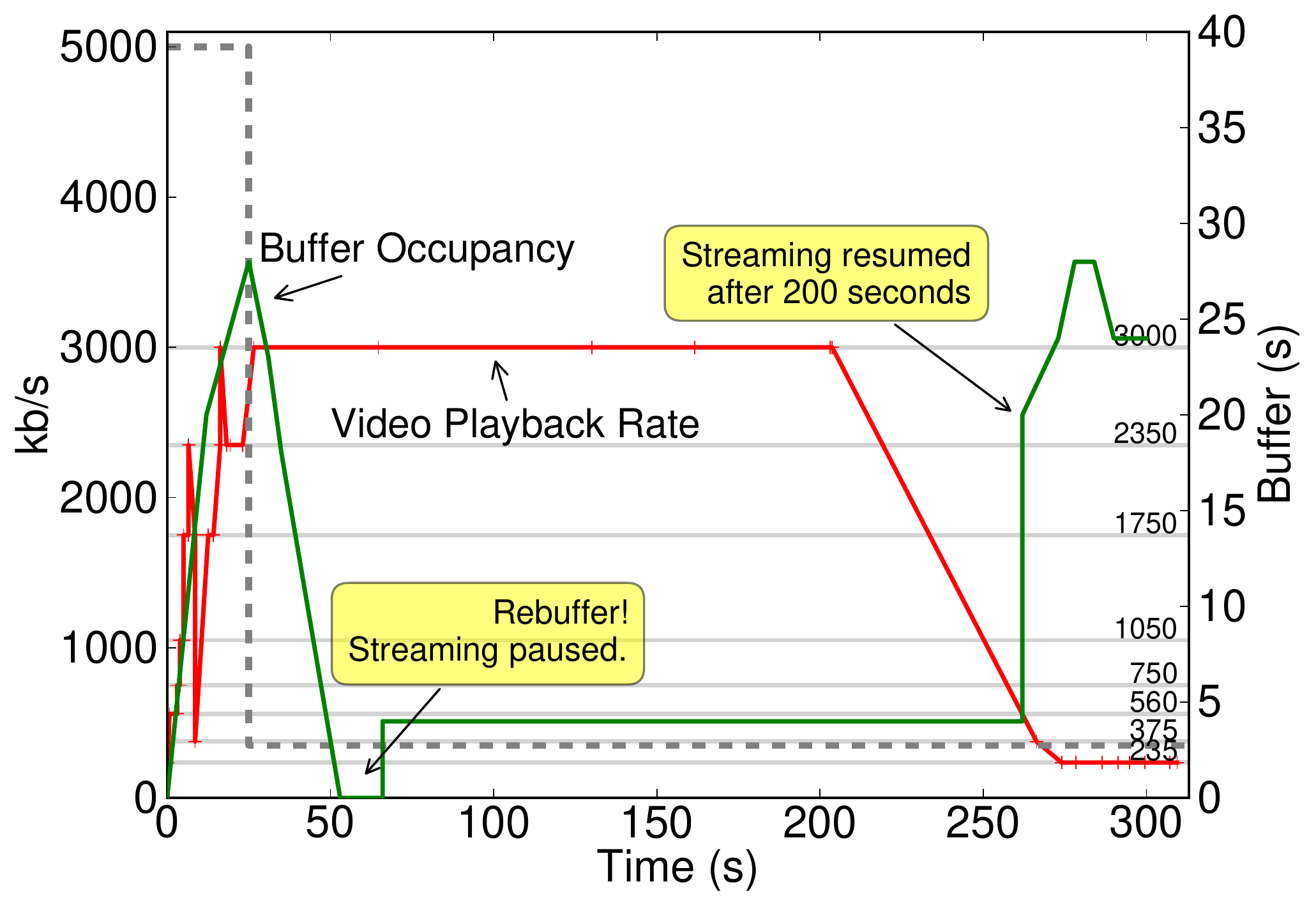}
\caption{{\em Being too aggressive}: A video starts streaming at
  3Mb/s over a 5Mb/s network. After 25s the available
  capacity drops to 350 kb/s. Instead of switching down to a lower
  video rate, e.g., 235kb/s, the client keeps playing at 3Mb/s. As a
  result, the client rebuffers  and does not resume playing video for 200s.}
\label{fig:servicea_aggressive}
\end{figure}

However, we recently reported in~\cite{bufferbased_fhmn} the surprising finding that popular video streaming services frequently rebuffer unnecessarily, i.e., despite the fact that $C(t) > R_{\min}, \forall t$. Figure~\ref{fig:servicea_aggressive} from~\cite{bufferbased_fhmn}
shows an example where the video keeps playing at too high a rate after the capacity has dropped (the ABR algorithm is too aggressive).  
The client rebuffers and freezes playback for 200 seconds.  Rebuffering was unnecessary because $C(t) > R_{\min}$ over the entire time series. 

We start by asking: why do we observe unnecessary rebuffers?  Notice that the client in Figure \ref{fig:servicea_aggressive} needs to rebuffer because when the buffer occupancy was low, it should have moved to a {\em lower} video rate.  The main reason the client does not switch is that {\em it estimates the current capacity to be sufficient to sustain a higher video rate}.  As a result, despite the fact that capacity is sufficient to sustain $R_{\min}$, the client does not find its way to that video rate in time.

Based on this observation, we conjecture that current ABR algorithms perform suboptimally because {\em they are poor at estimating capacity $C(t)$}. Estimating the available capacity is fraught with difficulties because of complex interactions between the HTTP control loop, the ABR control loop and TCP congestion control. Ultimately, as in Figure \ref{fig:servicea_aggressive}, this can lead to unnecessary rebuffering.

Of course, as suggested above, one easy solution that eliminates unnecessary rebuffers is to simply stream at $R_{\min}$ all the time.  Thus it is insufficient to only minimize rebuffering: we need to consider how we can eliminate unnecessary rebuffering, while {\em simultaneously} delivering high video quality to the user. This leads to the natural question: If direct estimation of available capacity is error-prone, then on what basis should an ABR algorithm pick the video rate?

Our informal discussion above suggests a path forward.  The easiest way to ensure that the algorithm never unnecessarily rebuffers is to simply request rate $R_{\min}$ when the buffer approaches empty, so that the buffer starts growing again as long as $C(t) > R_{\min}$.  
As the buffer grows, it is safe to increase $R(t)$ up to the maximum video rate as a full buffer is approached.  This suggests our solution: {\em an ABR algorithm that picks the video rate as a function of the playback buffer occupancy}.

\begin{figure}
\centering
\includegraphics[width=0.65\columnwidth]{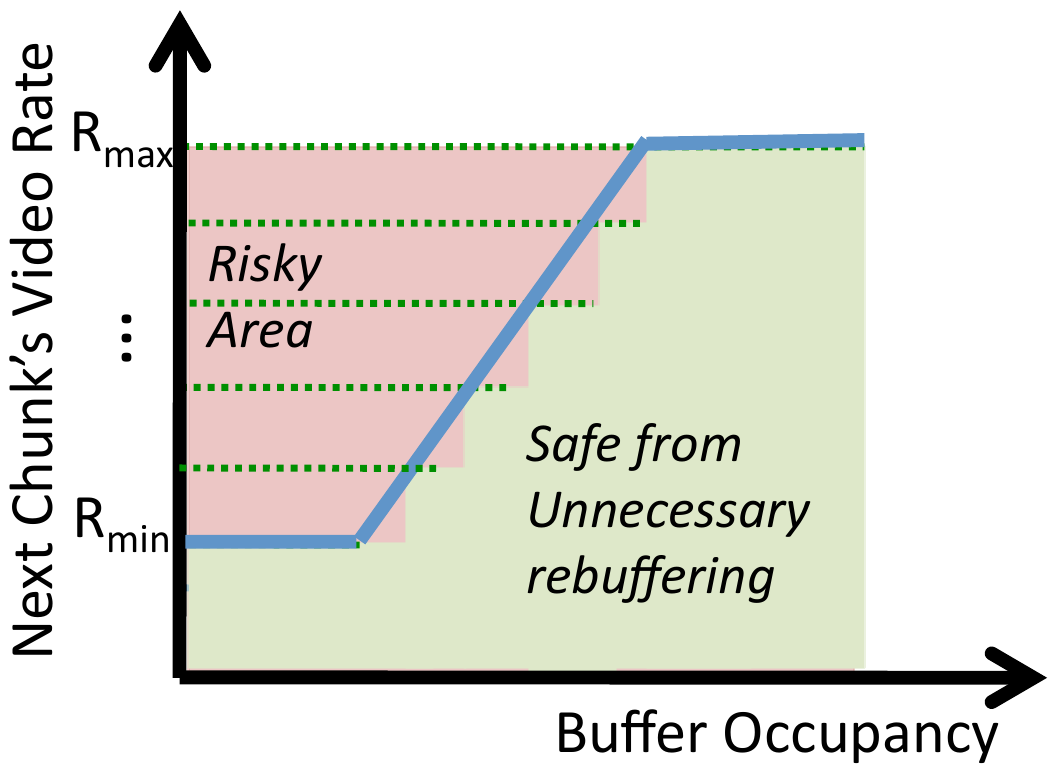}
\caption{Video rate as a function of buffer occupancy.}
\label{fig:ratemap_concept}
\end{figure}

In this paper, we develop this intuitive idea and make it concrete. Specifically, {\em we propose a broad class of ABR algorithms that pick a video rate as a function of the current occupancy of the playback buffer}, as shown in Figure~\ref{fig:ratemap_concept}. With hindsight, it is perhaps not surprising that a buffer-based approach works well, because the occupancy of the playback buffer is the primary state variable we are trying to manage. We show -- both formally and by demonstration in a real video service -- that this class of algorithms avoids unnecessary rebuffers and yet can achieve a high average video rate.


In Section~\ref{sec:algo} we introduce the broad class of buffer-based algorithms (BBA) that never unnecessarily rebuffer. In section~\ref{sec:baseline}
we pick a simple baseline algorithm, called BBA-0, and test it with real Netflix users. Although very simple, the baseline reduces the rebuffer rate by 20\%
compared to existing ABR algorithms, but at the expense of a lower video rate.

We then set out to steadily improve upon the baseline algorithm by taking into account several real world factors. First, in Section~\ref{sec:vbr}, we take into account the fact that real streamed video is encoded at a variable bit-rate (VBR); i.e. not all chunks at a given nominal rate are the same size. We refine the mapping function to correct for this, while maintaining the low rebuffer rate. Second, in Section~\ref{sec:start-up} we show that it helps to partition video watching into two phases, the startup phase (the first few minutes of viewing) and the steady-state phase. For the startup phase, while the playback buffer is filling and contains less information, we explore how to enter the risky area in Figure~\ref{fig:ratemap_concept} and handle the tradeoff between ramping the video rate aggressively and avoiding unnecessary rebuffers. And third, we address other practical concerns, such as temporary network outages and frequent rate switching. We provide mechanisms to tackle these issues in Section~\ref{sec:others}.

\section{The Class of Buffer-Based Algorithms}
\label{sec:algo}

\begin{figure}[t]
\centering
\includegraphics[width=0.8\linewidth]{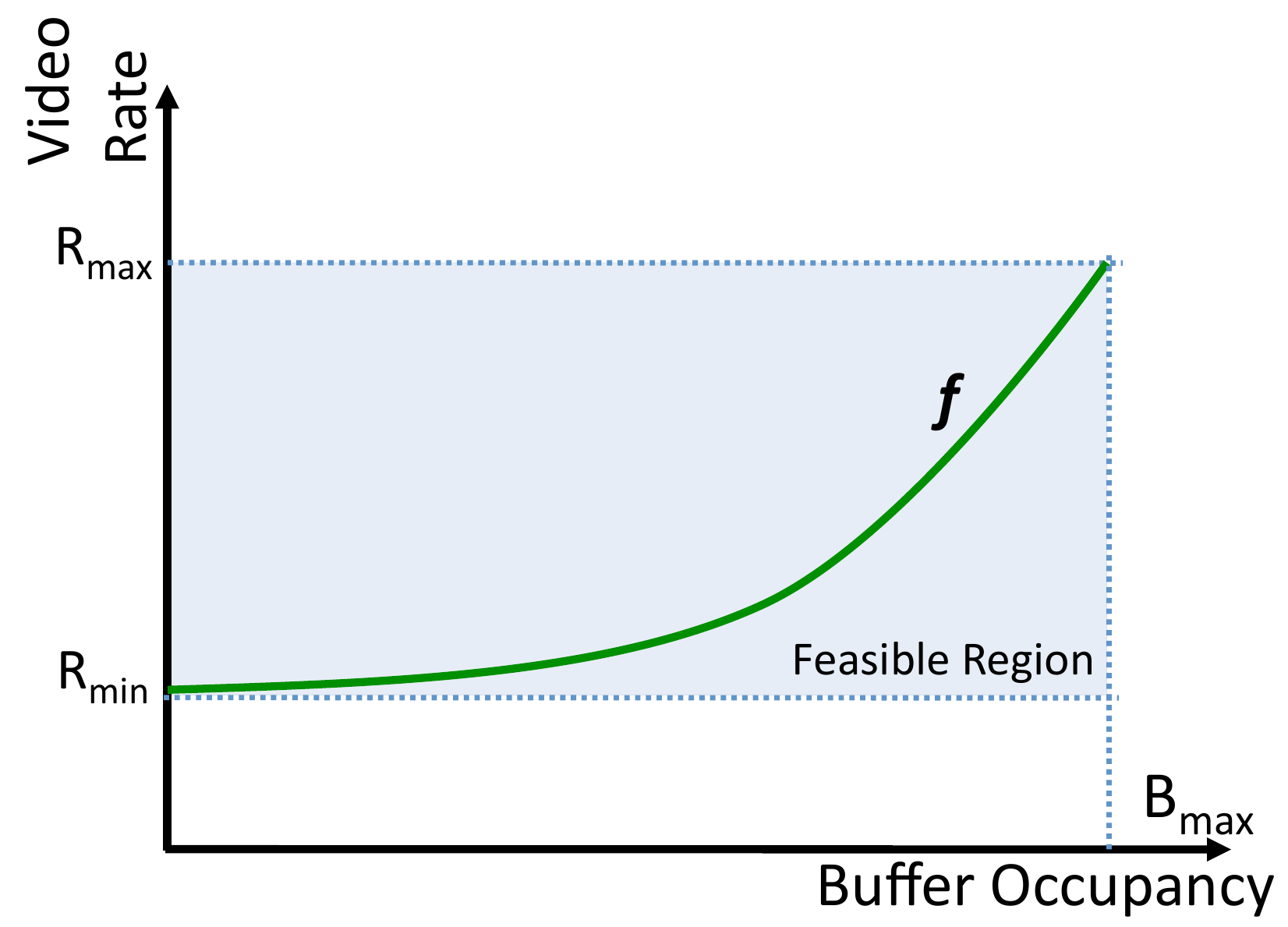}
\caption{The design space of rate maps.}
\label{fig:design_space}
\end{figure}

We say that an ABR algorithm is ``buffer-based'' if it picks the video rate as a function of the current buffer occupancy, $B(t)$. 
The design space for this class of algorithms can be expressed as the
buffer-rate plane shown in Figure~\ref{fig:design_space}.  
The shaded region between $[0, B_{\max}]$ on the buffer-axis and $[R_{\min}, R_{\max}]$
on the rate-axis defines the feasible region.  Any curve $f(B)$ on the
plane within the feasible region defines a {\em rate map}, a function that produces a video rate between $R_{\min}$ and $R_{\max}$ given the
current buffer occupancy.  We focus on rate maps that are {\em
continuous} functions of the buffer occupancy $B$, and that are {\em
strictly increasing} in the region $\{ B : R_{\min} < f(B) <
R_{\max} \}$.

In the following, we will only consider rate maps $f$ that are pinned at both ends: $f(0) = R_{\min}$ and
$f(B_{\max}) = R_{\max}$, as shown in Fig.~\ref{fig:design_space}.  
In other words, the rate map moves from the
lowest to highest video rate as the buffer moves from empty to full.
In~\cite{bufferbased_fhmn} we formally proved that 
any such rate map automatically meets our design principle. We summarize the proof here:

{\bf No unnecessary rebuffering:} As long as $C(t) \geq R_{\min}$ for all $t$ and we adapt $f(B) \to R_{\min}$ as $B \to 0$, 
we will never unnecessarily rebuffer because the buffer will start to grow before it runs dry.

{\bf Average video rate maximization:} As long as $f(B)$ is (1) increasing and (2) eventually reaches $R_{\max}$,
the average video rate will equal to the average capacity when $R_{\min} < C(t) < R_{\max}$ for all $t > 0$.

\section{The BBA-0 Algorithm}
\label{sec:baseline}

\begin{figure}[t]
\centering
\includegraphics[width=0.8\linewidth]{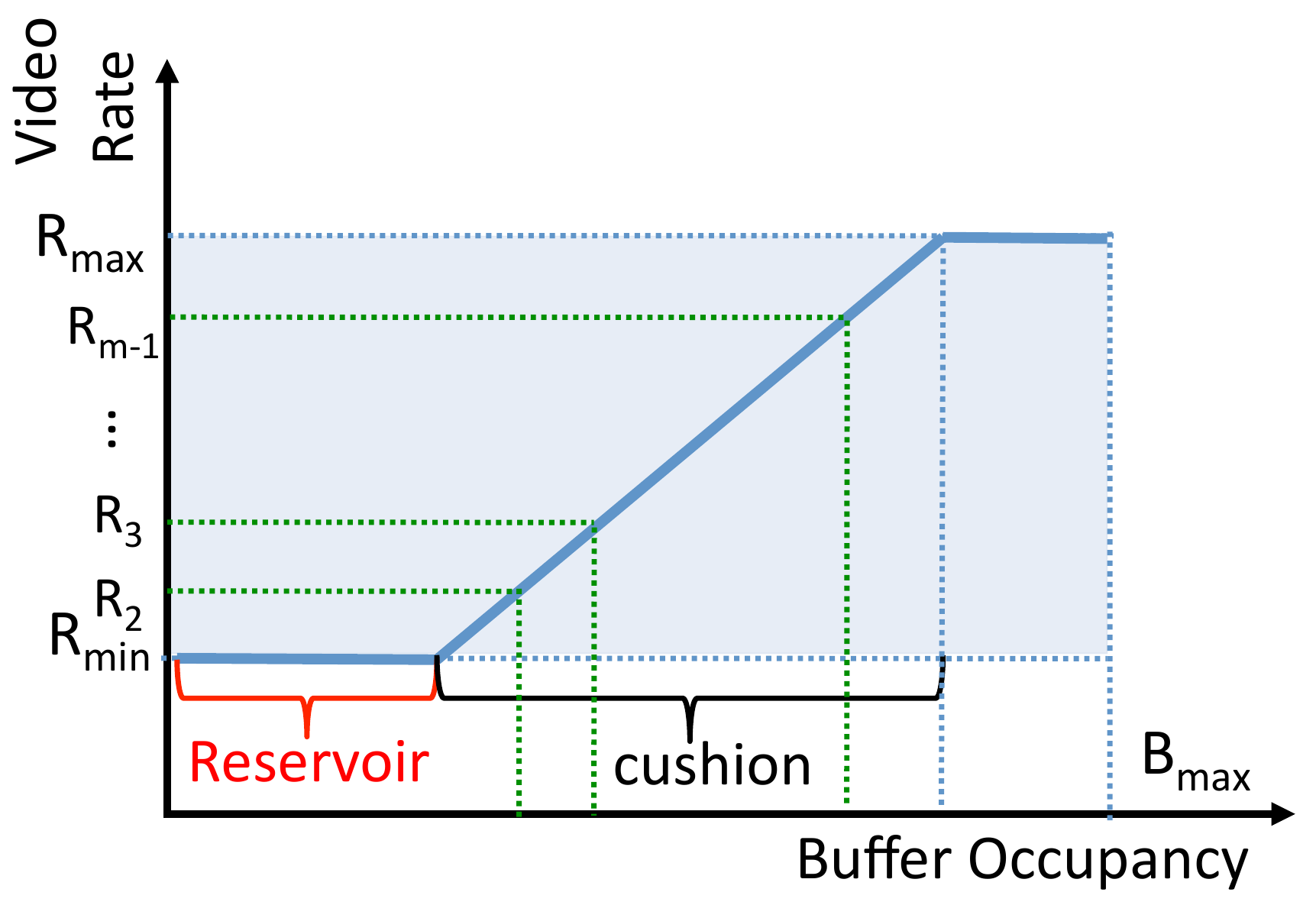}
\caption{The rate map used in the BBA-0 buffer-based algorithm.}
\label{fig:baseline_ratemap}
\end{figure}

The buffer-based algorithm used in our first round of experiments operates over a rate map $f$ pinned at both ends, i.e., $f(0) = R_{\min}$ and $f(B_{\max}) = R_{\max}$. In practice, because video arrives in chunks and we only pick a new rate when a chunk {\em finishes} arriving, we should not wait until the buffer is very close to empty before switching down to $R_{\min}$. If the network suddenly gets slower while we are in the middle of downloading a big chunk, the buffer could run dry before we even get a chance to switch to a lower rate. We therefore shift the rate map to the right so we switch to $R_{\min}$ sooner, creating an extra {\em reservoir}.

We discuss how to calculate the size of the reservoir more intelligently in Section~\ref{sec:vbr}. To start out simply, and to test that our buffer-based algorithms work in practice, we simply shift the rate map by 90s (a randomly picked value), as shown in Figure~\ref{fig:baseline_ratemap}. When $0 \leq B(t) \leq 90$, it will request video rate $R_{\min}$. Once the reservoir is full, we linearly increase the video rate until the rate map reaches $R_{\max}$.  
We call the above algorithm {\em BBA-0} in the following, since it is the simplest of our buffer-based algorithms.

Note that a rate map by itself does not fully define the algorithm.  Since the rate map
is continuous, it may not directly correspond to an available discrete video rate, $R_{\min}, R_2, R_3 ... R_{m-1}, R_{\max}$.

We therefore use the rate adaptation algorithm described in 
Algorithm~\ref{algo:nextRate}. The algorithm follows a
simple principle: it stays at the current video rate as long
as the rate suggested by the rate map does not pass either the next
higher available video rate ($\rate_{+}$) or the next lower available
video rate ($\rate_{-}$).
If either of these ``barriers'' are hit, the rate is switched up or
down (respectively) to a new discrete value suggested by the rate map.
In this way, the buffer distance between the adjacent
video rates provides a natural cushion to absorb rate oscillations, making the video rate a little ``sticky''.

\begin{algorithm}
\caption{Video Rate Adaptation Algorithm}
\label{algo:nextRate}
\SetAlgoLined
\KwIn{ $\rate_{\prev}$: The previously used video rate \\ \quad\quad\quad\quad $\buf_{\now}$: The current buffer occupancy}
\KwOut{ $\rate_{\nxt}$: The next video rate }
\BlankLine
\uIf{$\rate_{\prev} = R_{\max}$}{
$\rate_{+} = R_{\max}$
}
\uElse{
$\rate_{+} = \min \{ R_i : R_i > \rate_{\prev} \}$
}

\uIf{$\rate_{\prev} = R_{\min}$}{
$\rate_{-} = R_{\min}$
}
\uElse{
$\rate_{-} = \max \{ R_i : R_i < \rate_{\prev} \}$
}

\uIf{$f(\buf_{\now}) \geq \rate_{+}$}{
    $\rate_{\nxt} = \max \{ R_i : R_i < f(\buf_{\now}) \}$\;
}
\uElseIf{$f(\buf_{\now}) \leq \rate_{-}$}{
    $\rate_{\nxt} = \min \{ R_i : R_i > f(\buf_{\now}) \}$\;
}
\uElse{
    $\rate_{\nxt}$ = $\rate_{\prev}$\;
}
\KwRet{$\rate_{\nxt}$}\;
\end{algorithm}

\begin{figure}[t]
\centering
\includegraphics[width=\linewidth]{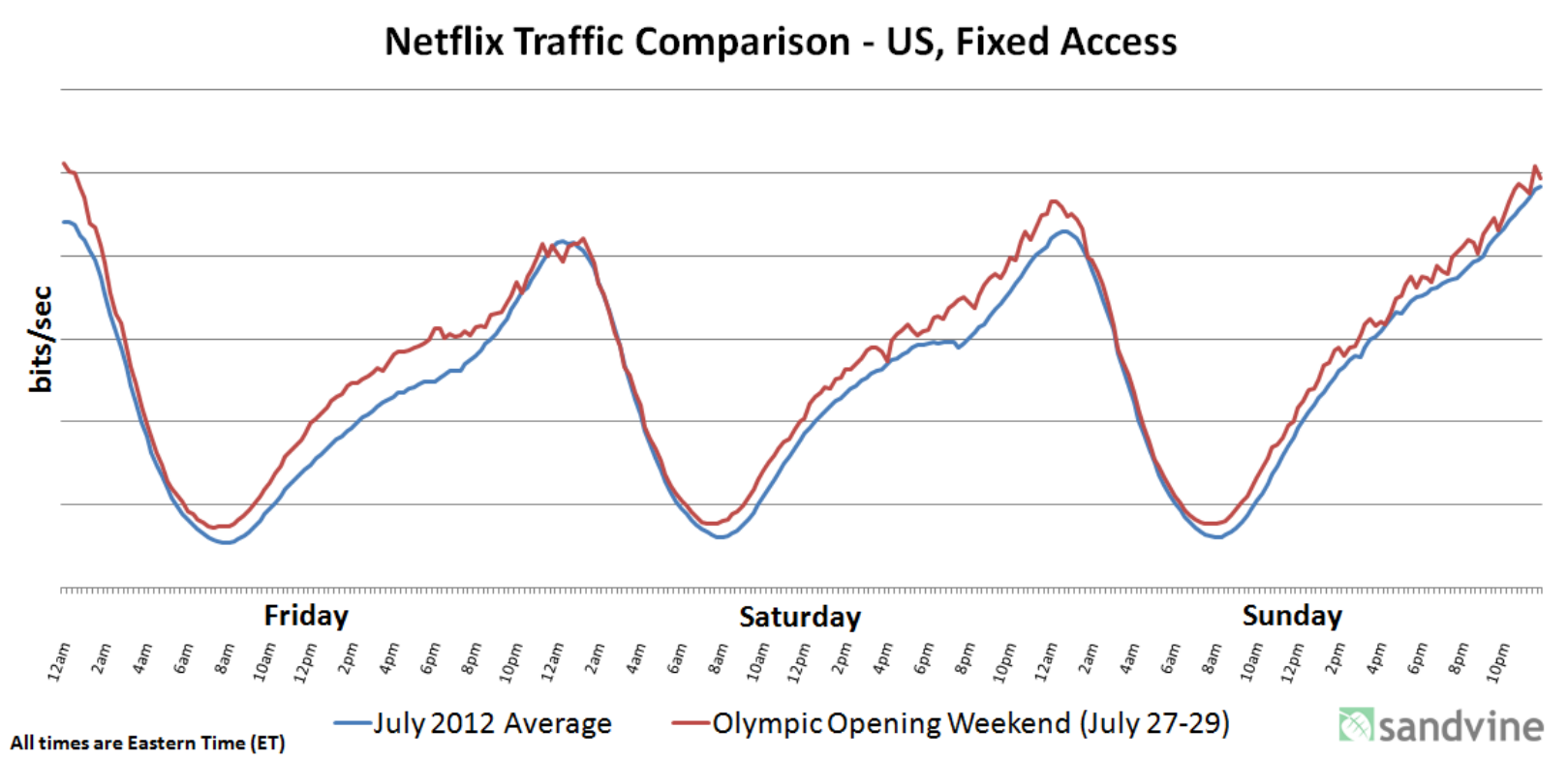}
\caption{A public report (compiled by Sandvine~\cite{sandvine-report}) showing the diurnal traffic pattern in the Netflix video service. The two lines both represent Netflix traffic in July 2012 before and during the London Olympics. Note that the traffic peaks during 8pm-1am EDT (0-5am GMT).}
\label{fig:sandvine}
\end{figure}

\subsection{Experiments}
We implemented the BBA-0 algorithm in one of the Netflix video clients.\footnote{Netflix is the largest video streaming service with over 36 million users.} The video client is browser-based with a 240s playback buffer, and downloads the ABR algorithm at the start of the video session. We selected a portion of users around the world to take part in the experiments over the weekend of September 6th to 9th, 2013. We divided them equally into three groups, each with a different ABR algorithm. 

The first group of users is our control group who use the Netflix ABR algorithm.\footnote{This was the Netflix algorithm at the time of our experiments. The Netflix algorithm keeps evolving, and so the current algorithm is different.} We call this the ``control'' algorithm. The control algorithm has been improved over the past five years to perform well under many conditions. It picks a video rate primarily based on capacity estimation, but also uses buffer occupancy as a secondary signal. It is considered one of the best-performing algorithms in the industry.

The second group of users always stream at $R_{\min}$, and we call this degenerate algorithm ``$R_{\min}$ Always''.  Always operating at the lowest video rate minimizes the chances of the buffer running dry, giving us a lower bound on the rebuffering rate to compare new algorithms against. For most sessions $R_{\min} = 560kb/s$, but in some cases it is $235kb/s$.\footnote{In Netflix, $R_{min}$ is normally $235$ kb/s. However, most customers can sustain 560kb/s, especially in Europe. If a user historically sustained 560kb/s we artificially set $R_{min} = 560kb/s$ to avoid degrading the video experience too far. The mechanism to pick $R_{\min}$ is the same across all three test groups.}

The third group of users use our new BBA-0 algorithm.

All three user groups are distributed similarly over ISPs, geolocations, watching behaviors and devices. The only difference between the three groups of clients is the rate selection algorithm; they share the same code base for other mechanisms, such as CDN selections and error handling. In total, each group of users consumed roughly $120,000$ viewing hours. 
Since the video streaming peaks between 0-5am GMT, as shown in Figure~\ref{fig:sandvine}, in the rest of the paper we refer to 0-5am GMT as the {\em peak} period and the rest of a day as the {\em off-peak} period. To evaluate the performance, we will compare the overall number of rebuffers per playhour and the average delivered video rate in each group.


\subsection{Results}

\begin{figure}[t]
\centering
\includegraphics[width=\linewidth]{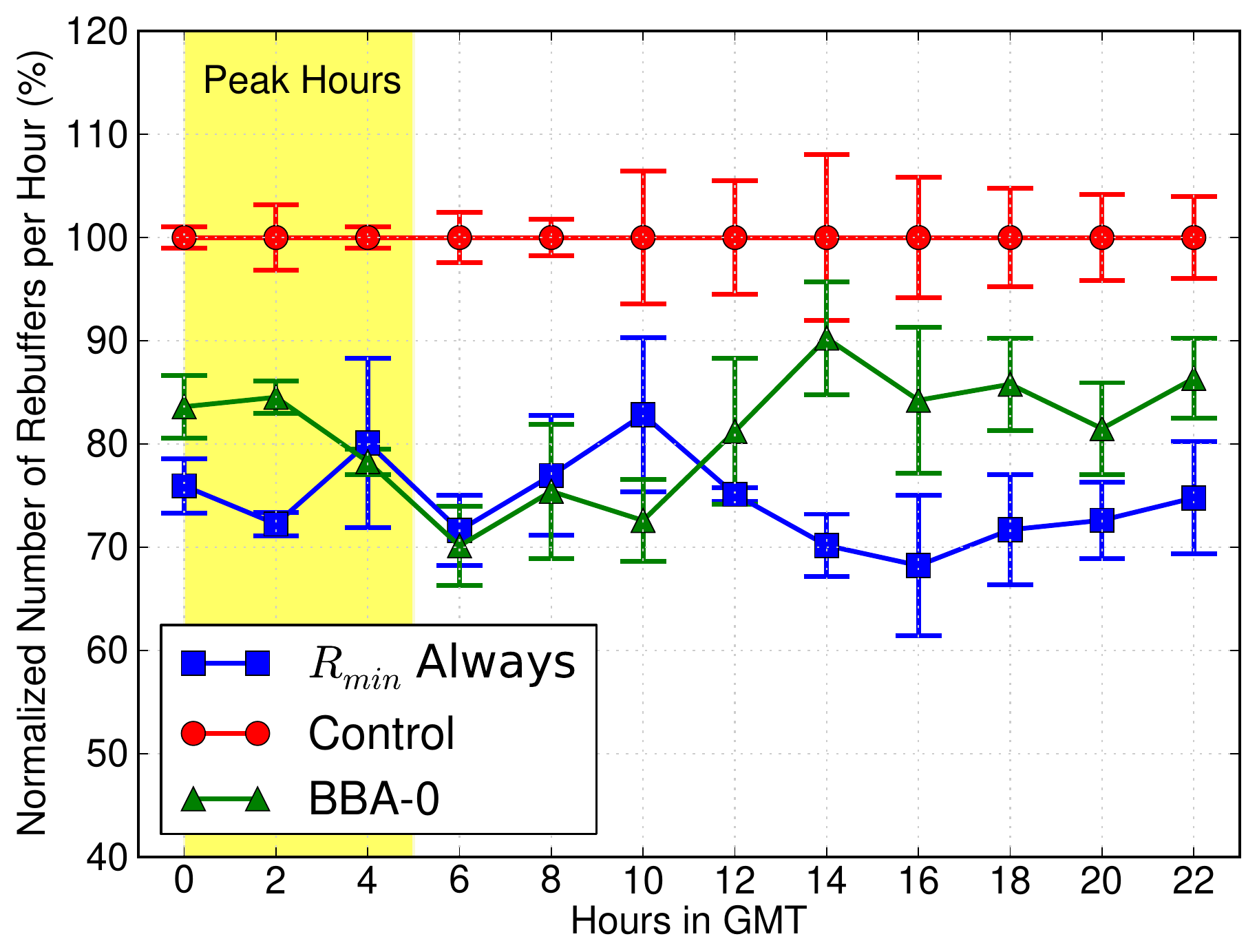}
\caption{Number of rebuffers per playhour for the {\em Control},  {\em $R_{\min}$ Always} and {\em BBA-0} algorithms. The numbers are normalized to the average rebuffer rate of the control group in each two hour period.}
\label{fig:guinea_rebuf}
\end{figure}



\noindent{\bf Rebuffer Rate.}
Figure~\ref{fig:guinea_rebuf} plots the number of rebuffers per play-hour throughout the day. Note the numbers are normalized to the average rebuffer rate of the {\em control} group in each two-hour period. The first thing to notice is that {\em $R_{\min}$ Always} and BBA-0 always have a lower rebuffer rate than the {\em control} algorithm.  The difference between the {\em control} algorithm and the {\em $R_{\min}$ Always} algorithm suggests that 20 - 30\% of the rebuffers might be caused by poor choice of video rate.

During the middle-of-night period in the USA just after peak viewing (6am-12pm GMT), BBA-0 matches the {\em $R_{\min}$ Always} lower bound very closely. This is because the viewing rate is relatively low (see Figure~\ref{fig:sandvine}) and the overall Internet usage is low, and so the network capacity does not change much. During peak hours, BBA-0 is close to, but slightly worse than, the {\em $R_{\min}$ Always} algorithm. However, the rebuffer rate is further from the lower bound during the rest of the day. Still, the BBA-0 algorithm consistently has a 16 - 29\% lower rebuffer rate than the {\em control} algorithm. This is encouraging given the extremely simple nature of the BBA-0 algorithm. Still, we hope to do better. In the following sections we try to improve BBA-0. 


\begin{figure}[t]
\centering
\includegraphics[width=\linewidth]{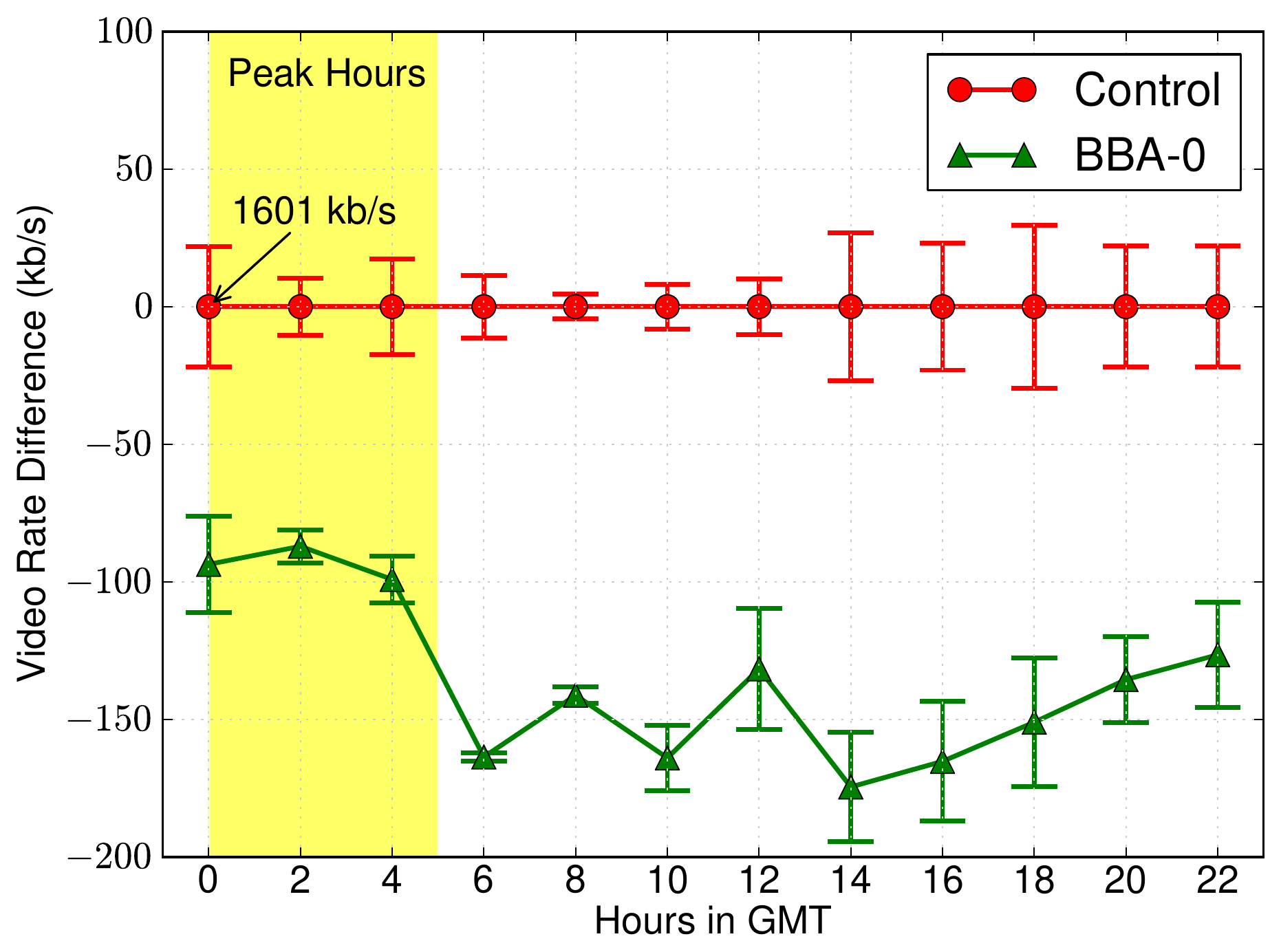}
\caption{The difference on video rate per two hour window between the BBA-0 algorithm and the control.}
\label{fig:guinea_vrate}
\end{figure}

\noindent{\bf Video Rate.} Figure~\ref{fig:guinea_vrate} shows the difference in the delivered video rate between the BBA-0 algorithm and the control algorithm. 
Since ``$R_{\min}$ Always'' always streams at $R_{\min}$ (except when rebuffering), its delivered video rate is a flat line and is excluded from the figure. 
The BBA-0 algorithm is roughly 100 kb/s worse than the {\em control} algorithm during peak hours, and 150 kb/s worse during off-peak. There are two main reasons for the degradation in video quality. First, our BBA-0 algorithm assumes video is encoded at a constant rate (CBR) when in fact it is VBR. Second, and more significantly, while the reservoir is filling during the start-up period, our BBA-0 algorithm always requests video at rate $R_{\min}$. Given we picked a 90s reservoir, it takes a long time to fill the buffer and is a non-negligible fraction of the average session length. We address both issues in the following two sections. 


\begin{figure}[t]
\centering
\includegraphics[width=\linewidth]{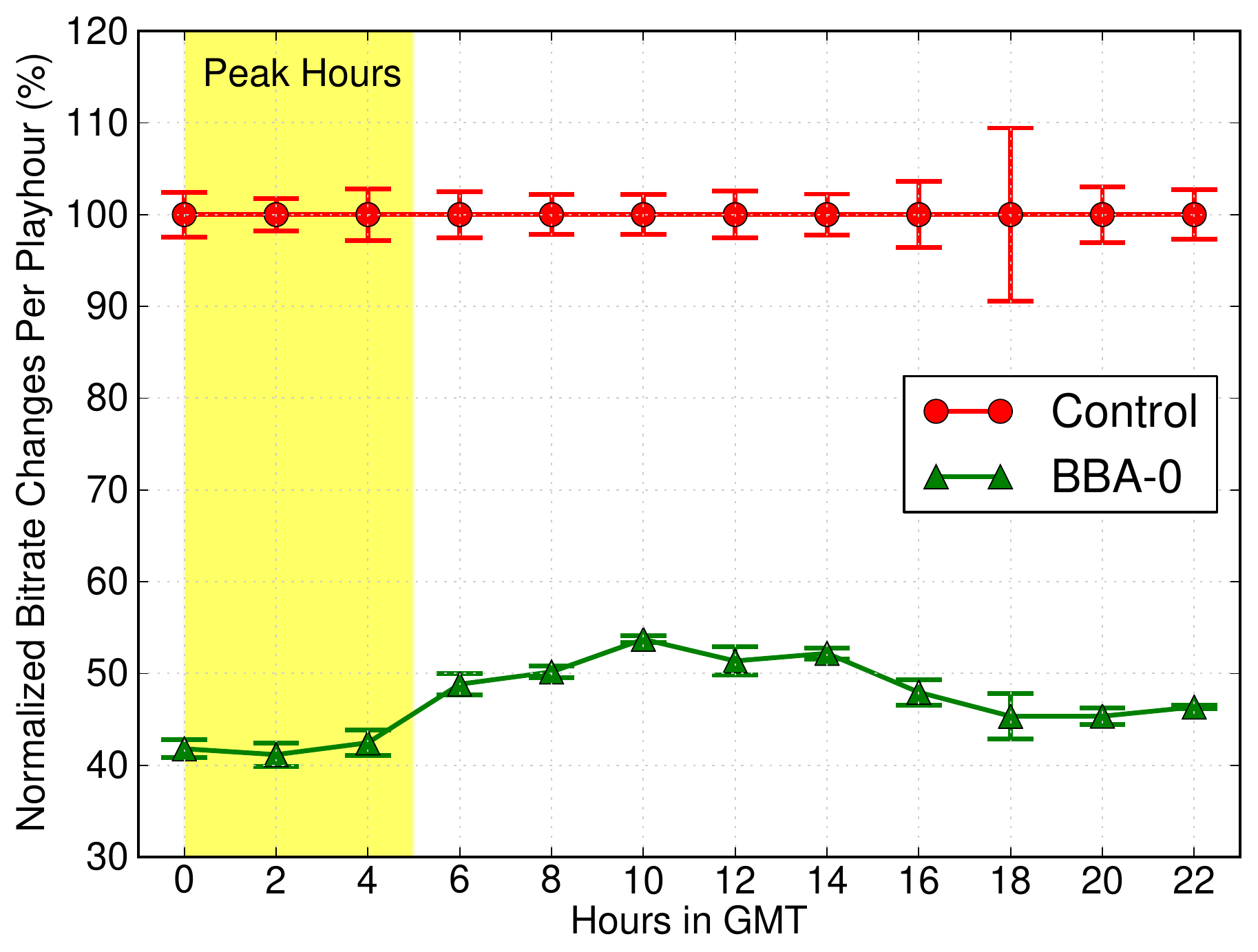}
\caption{Average video switching rate per two hour window for the {\em Control} and {\em BBA-0} algorithms. The numbers are normalized to the average switching rate of the control group for each window.}
\label{fig:guinea_crate}
\end{figure}

\noindent{\bf Video Switching Rate. } Since our BBA-0 algorithm picks the video rate based on the buffer level, we can expect the rate to fluctuate as the buffer occupancy changes. However, Algorithm~\ref{algo:nextRate} uses the distance between adjacent video rates to naturally cushion, and absorb, rate oscillations.
Figure~\ref{fig:guinea_crate} compares BBA-0 with the {\em control} algorithm. 
Note the numbers are normalized to the average switching rate of the control group for each two-hour period. 
The BBA-0 algorithm reduces the switching rate by roughly 60\% during peak hours, and by roughly 50\% during off-peak hours.

\section{Handling Variable Bitrate (VBR)}
\label{sec:vbr}

In Section~\ref{sec:baseline}, the BBA-0 algorithm treats videos as constant bit rate (CBR) and randomly sets the reservoir size to a large value. 
Although we are able to get a significant amount of reduction in rebuffering compare to the control,
there is still room to improve when comparing to the ``$R_{\min}$ always'' line. 
Also, the video rate achieved by the BBA-0 algorithm is significantly lower than the control.
In this section, we will discuss mechanisms to further improve on both rebuffer rate and video rate by taking the encoding scheme into consideration.


\begin{figure}
\centering
\includegraphics[width=\linewidth]{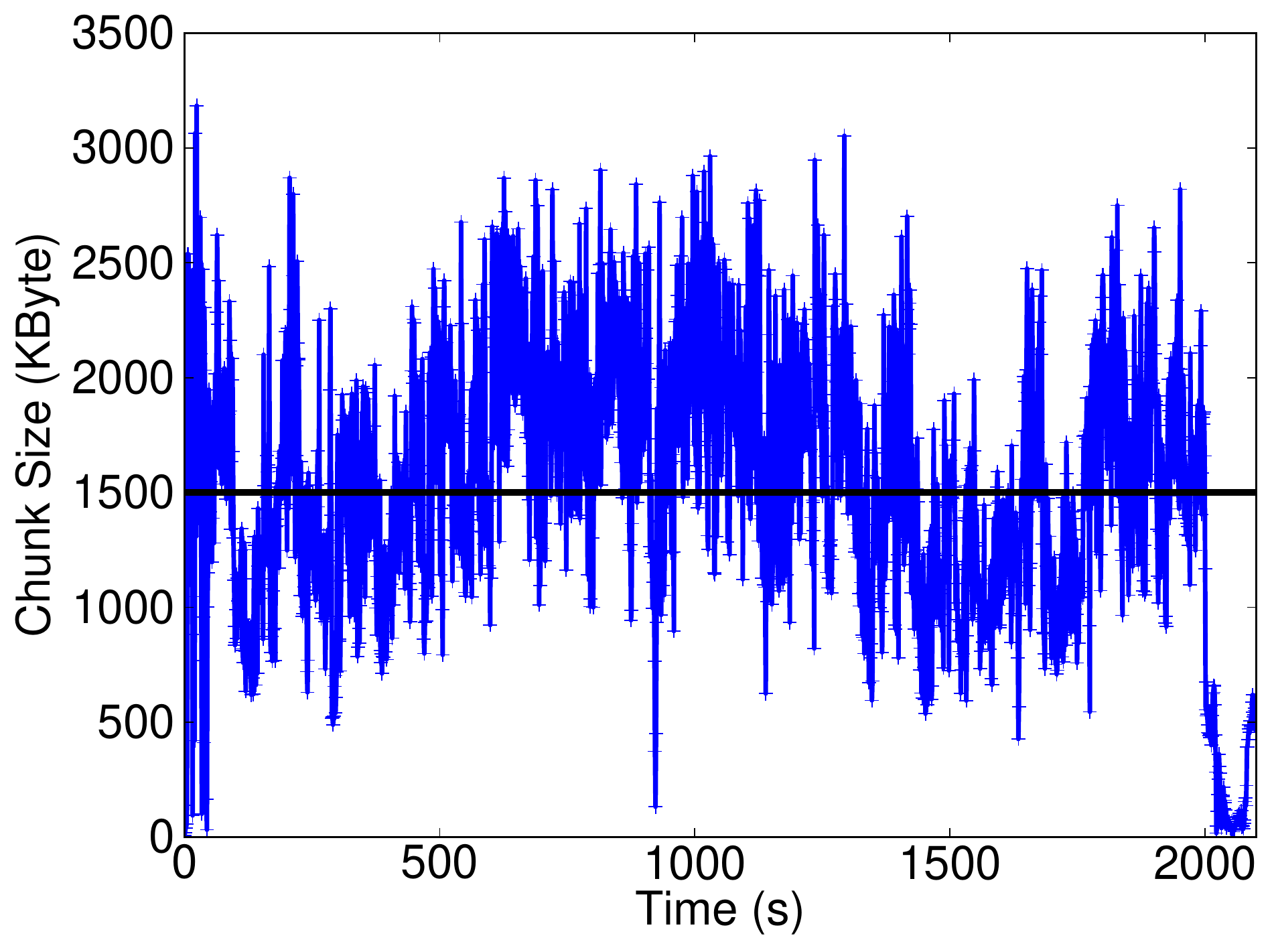}
\caption{The size of 4-second chunks of a video encoded at an average rate of 3Mb/s.}
\label{fig:vbrexample}
\end{figure}

In practice, most of the video streaming services encode their videos in variable bitrate (VBR),
  since it allows more flexibility and can use bytes more efficiently.
When a video is encoded in VBR at a nominal video rate, the nominal rate represents the {\em average} video rate 
and the instantaneous video rate varies around the average value.
As a result, the chunk size will not be uniformly identical in a stream of a given rate.
Figure~\ref{fig:vbrexample} shows the size of $4$-second chunks over time from a Netflix video encoded at $3$ Mb/s.
The black line represents the average chunk size. 
As we can see from the figure, the variation on chunk size can be quite significant within a single video rate.

\begin{figure}[t]
\centering
\includegraphics[width=\linewidth]{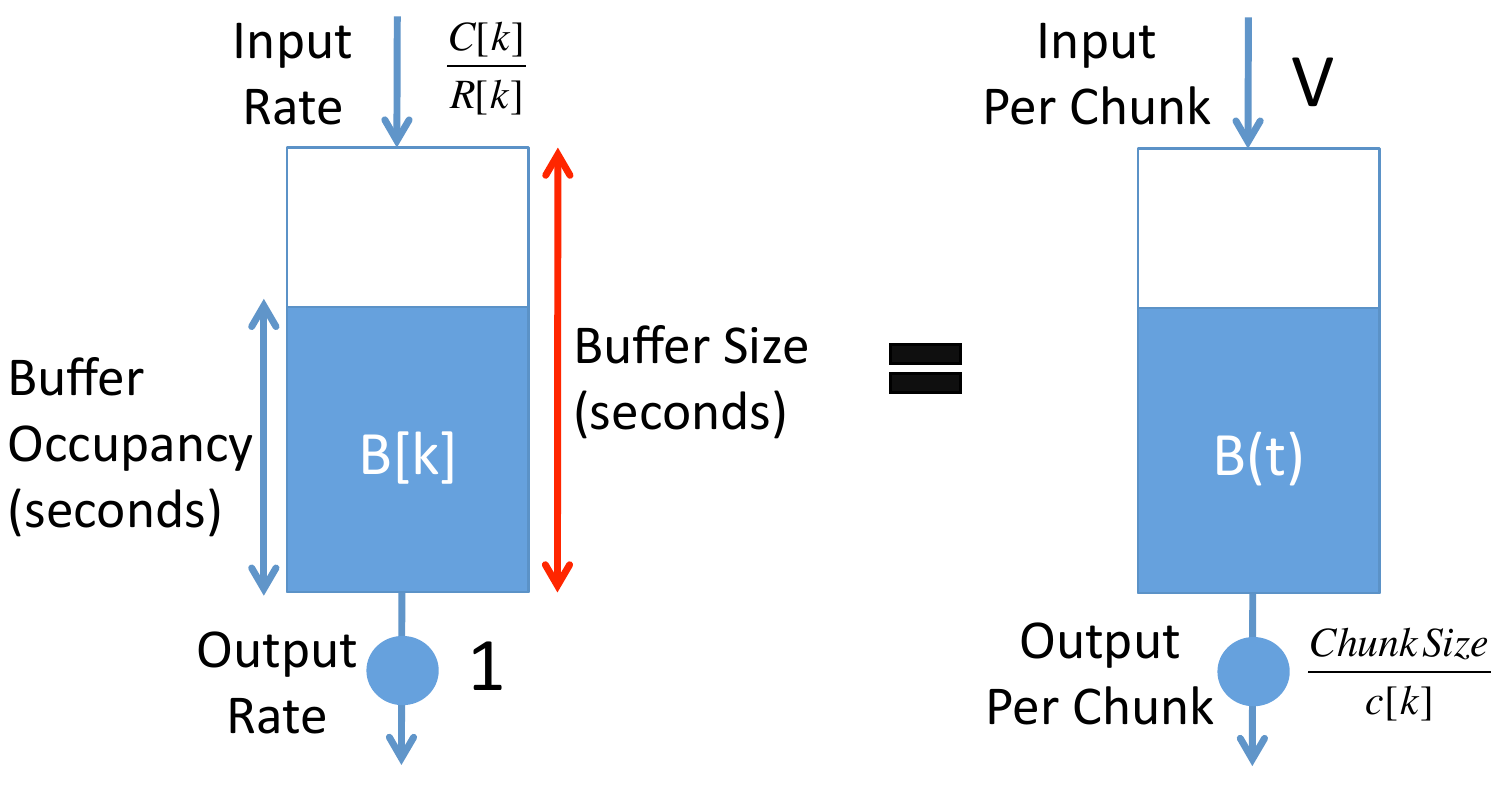}
\caption{Two equivalent models of the streaming playback buffer.}
\label{fig:buffer_model}
\end{figure}

Let's now re-consider the buffer dynamics under VBR. 
Given the constraint that we can only select video rates on a
chunk-by-chunk basis, it is useful to consider the buffer dynamics
when observed {\em at the time points when a chunk finishes}, 
as shown in Figure~\ref{fig:buffer_model}.
Let $r[k]$ be the video rate selected for the $k$-th chunk 
and $c[k]$ be the average system capacity during the download of the $k$-th chunk.
For the $k$-th chunk from the stream of video rate $r$, we denote the chunk size as $Chunk[r][k]$. 
Since each chunk still contains $V$ seconds of video, the buffer now drains $Chunk[r][k]/c[k]$ seconds while it fills with $V$ seconds of video. 

\subsection{Reservoir Calculation}

\begin{figure}
\centering
\includegraphics[width=\linewidth]{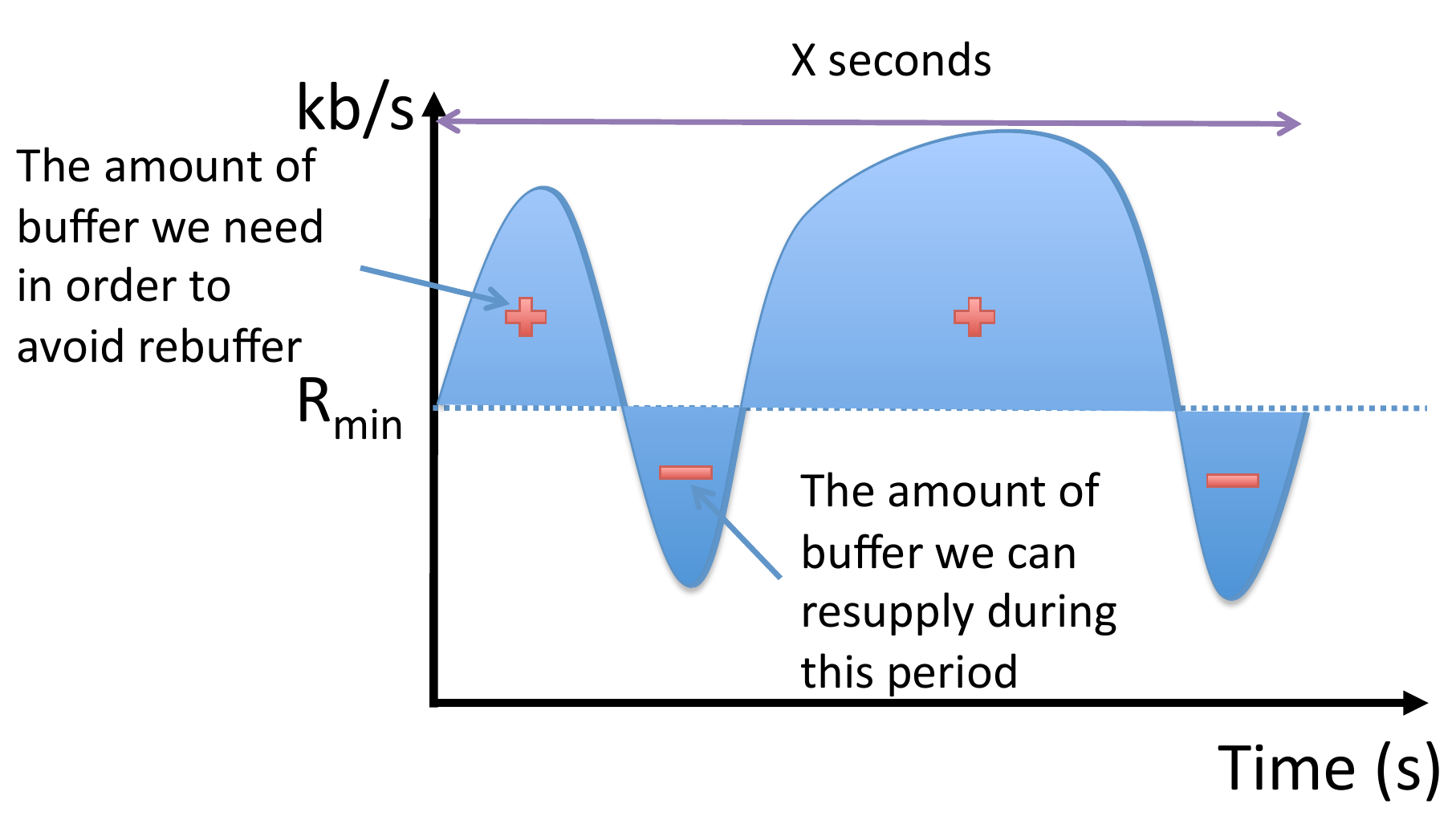}
\caption{The size of reservoir is calculated from the chunk size variation.}
\label{fig:reservoir_calc}
\end{figure}


Since the instantaneous video rate can be much higher than the nominal rate in VBR, we will need an extra reservoir to absorb the buffer variation caused by the chunk size variation.
The size of reservoir should be big enough to ensure the client can continue playing at $R_{\min}$ even when the capacity $c[k]$ is exactly equal to $R_{\min}$.
Assuming $c[k] = R_{\min}$, the video client will consume more video in the buffer than the input when the chunk size is larger than the average, $V R_{\min}$. 
On the other hand, when the chunk size is lower than $V R_{\min}$, the buffer is consumed more slowly than the input and the buffer occupancy will increase. 
Thus, by summing up the amount of buffer the client will consume minus the amount it can resupply during the next $X$ seconds, 
we can figure out the amount of reservoir we need.
Figure~\ref{fig:reservoir_calc} summarizes how the calculation is done. 
In the implementation, we set $X$ as twice of the buffer size, i.e., $480$ seconds. 
As a practical matter, since the calculated reservoir size depends highly on the specific video, we further bound the size of reservoir to be between 8 seconds to 140 seconds.

\subsection{Chunk Map}

\begin{figure}[t]
\centering
\includegraphics[width=\linewidth]{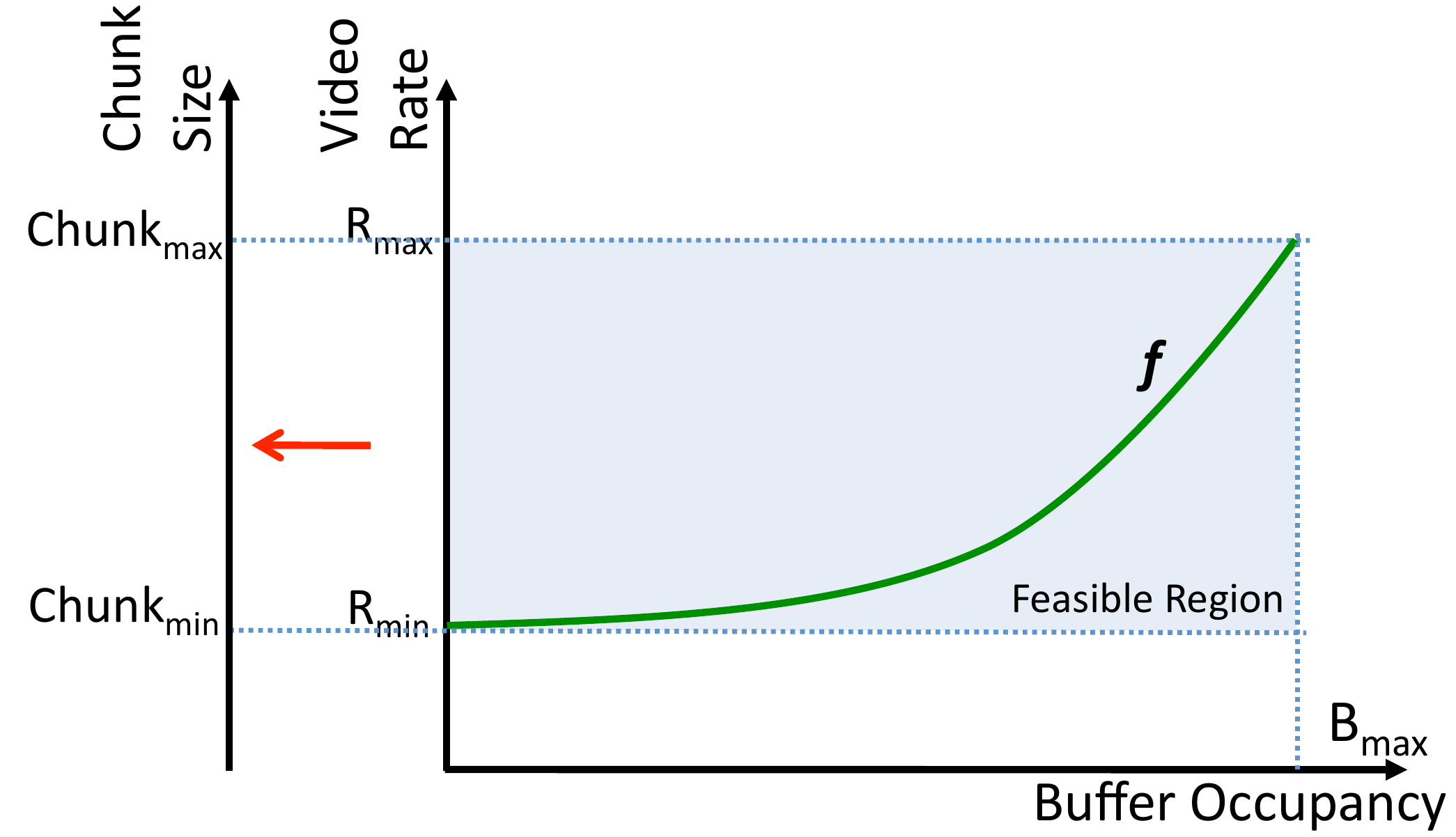}
\caption{To handle VBR, we generalize the concept of {\em rate map} to {\em chunk map} by transforming the Y-axis from video rates to chunk sizes.}
\label{fig:design_space_vbr}
\end{figure}

Since the buffer dynamics now depend on the chunk size of the upcoming video segments, instead of the video rate, it makes more sense to map the buffer occupancy to the chunk size directly. 
In other words, we can generalize the design space and change it from the buffer-rate  plane to the buffer-chunk plane as shown in Figure~\ref{fig:design_space_vbr}.
Each curve in the figure now defines a {\em chunk map}, which represents the maximally allowable chunk size according to the buffer occupancy.
In the figure, the feasible region is now defined between $[0, B_{\max}]$ on the buffer-axis and $[Chunk_{\min}, Chunk_{\max}]$ on the chunk-axis, 
where $Chunk_{\min}$ and  $Chunk_{\max}$ represent the average chunk size in $R_{\min}$ and $R_{\max}$ respectively. 

We can now generalize Algorithm~\ref{algo:nextRate} to use the chunk map: it stays at the current video rate as long as the chunk size suggested by the map does not pass the size of the next upcoming chunk at the next higher available video rate ($\rate_{+}$) or the next lower available video rate ($\rate_{-}$). If either of these ``barriers'' are hit, the rate is switched up or
down (respectively). 
Note that by using the chunk map, we no longer have a fixed mapping between buffer levels and video rates. This could result in a higher frequency of video rate switches. We will address this issue in Section~\ref{sec:others}.

\subsection{Results}

We use the same setup as in Section~\ref{sec:baseline}.
We select the same number of users as in each group to use our VBR-enabled buffer-based algorithm, which dynamically calculates the size of reservoir and uses chunk map.
We will refer to the algorithm as {\em BBA-1} in the following, as it is our second iteration of the buffer-based algorithm. 
The experiment was conducted along with the experiment in Section~\ref{sec:baseline}, over the weekend of September 6th to 9th, 2013. 

\begin{figure}[t]
\centering
\includegraphics[width=\linewidth]{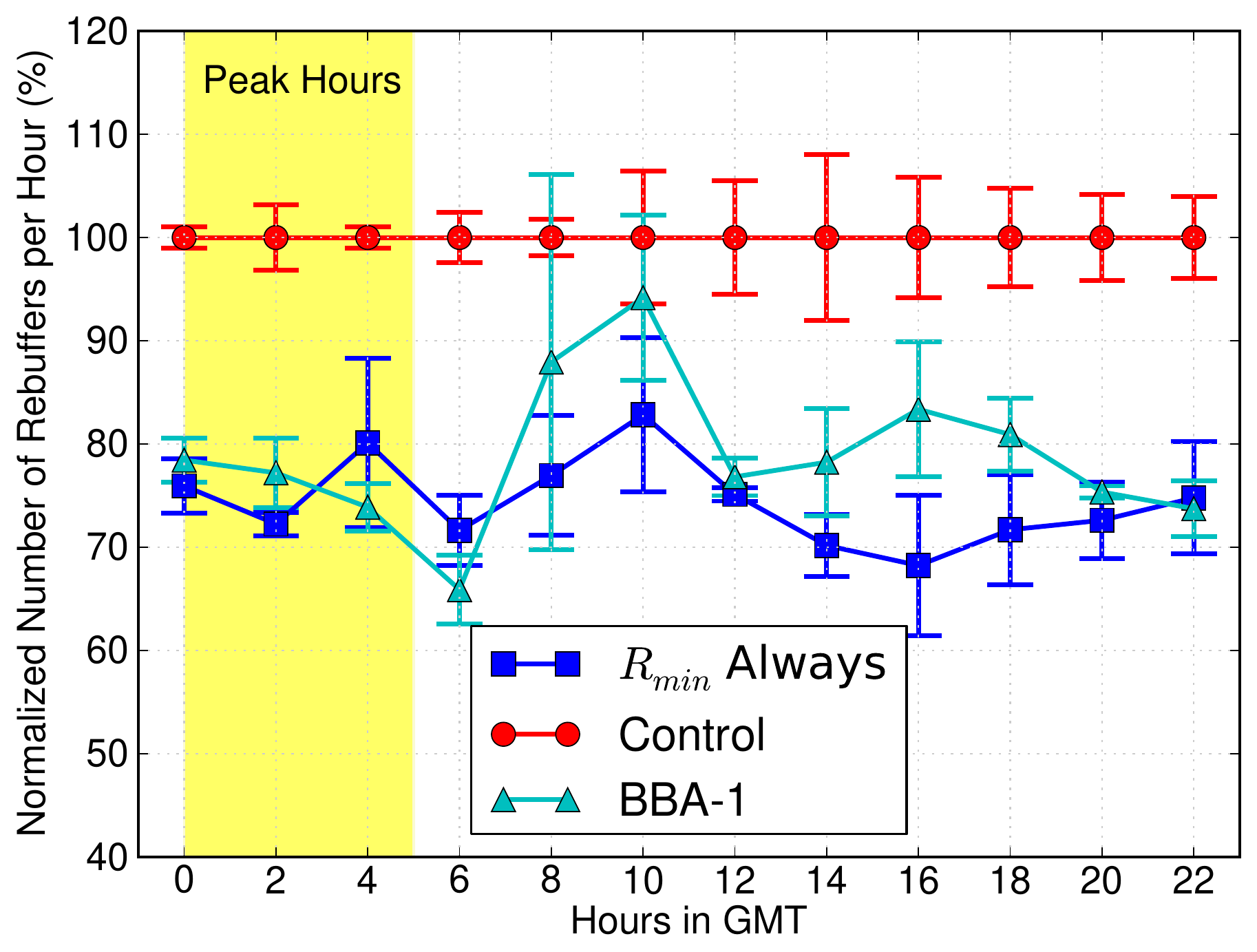}
\caption{The BBA-1 algorithm achieves close-to-optimal rebuffer rate, esepcially during the peak hours. Note the numbers are normalized to the average rebuffer rate of the control group for each window.}
\label{fig:rebuf_dr}
\end{figure}

Figure~\ref{fig:rebuf_dr} shows the rebuffer rate in terms of number of rebuffer per playhour. 
Note the numbers are again normalized to the average rebuffer rate of control in each window. 
We can see from the figure that the BBA-1 algorithm is very close to the optimal line and performs better than the BBA-0 algorithm. 
The improvement over the control is especially clear during the peak hours, where the BBA-1 algorithm provides 20 - 28\% of improvement on rebuffer rate.
During 8am to 10pm GMT, since the control algorithm has a very low rebuffer rate, the small difference between the three groups is magnified. 
The BBA-1 algorithm also improves the video rate compare to BBA-0 by 40 - 70 kb/s in average, but is still 50 - 120 kb/s away from the control algorithm. 
As we mentioned before, the discrepancy on video rate really comes from the start-up period.
If we compare the average video rate of the first 60 seconds between the BBA-1 algorithm and the control algorithm,
the BBA-1 algorithm achieves 700 kb/s less in average on video rate. 
This is because before the client builds up buffer to the size of reservoir, the BBA-1 algorithm will always request for $R_{\min}$, as it is the only safe rate given the buffer occupancy. 
If we want to further improve the video rate, we need to start taking some risk and enter into the ``risky'' area shown in Figure~\ref{fig:ratemap_concept}. 


\begin{figure}[t]
\centering
\includegraphics[width=\linewidth]{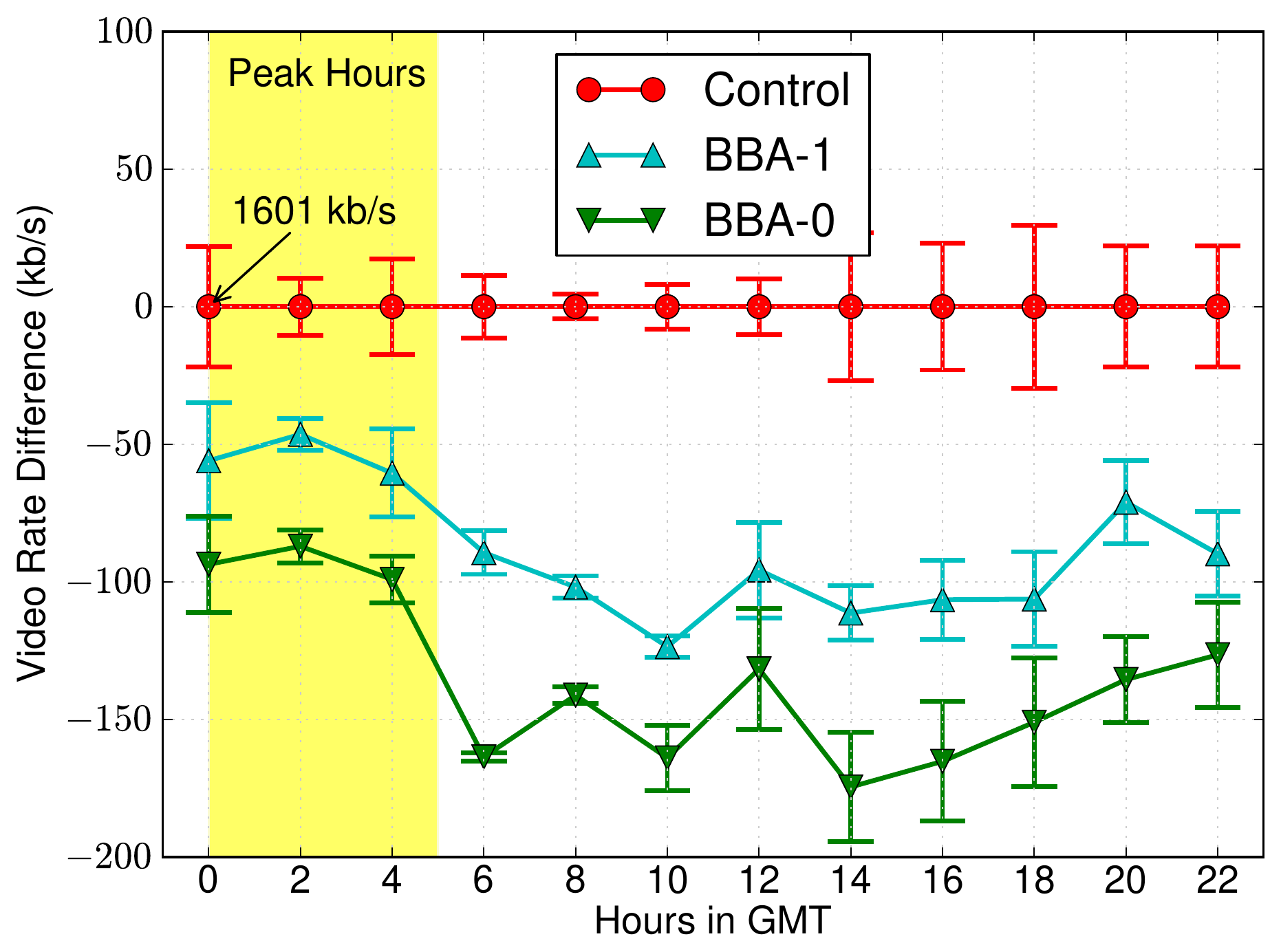}
\caption{The BBA-1 algorithm improved video rate by 40 - 70 kb/s compare to BBA-0, but still 50 - 120 kb/s away from the control.}
\label{fig:vrate_dr}
\end{figure}



\section{Boosting the video rate during start-up}
\label{sec:start-up}

\begin{figure}[t]
\centering
\includegraphics[width=\linewidth]{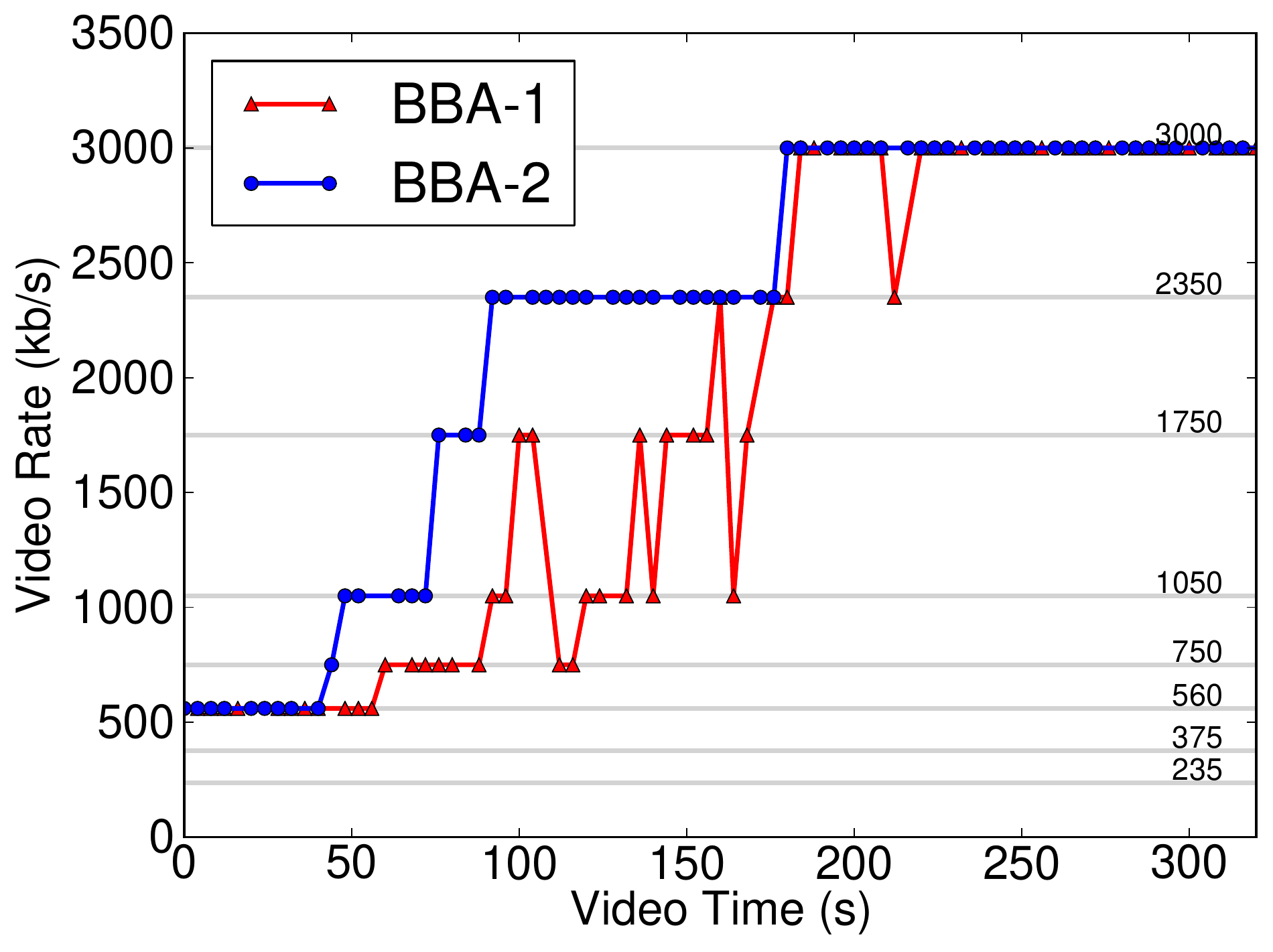}
\caption{Typical time series of video rates picked by BBA-1 (red) and BBA-2 (blue). BBA-1 ramps up slowly because it follows the chunk-map. BBA-2 ramps up faster to try and reach steady state sooner.}
\label{fig:startup_timeseries}
\end{figure}

Almost all of the difference in video rate between BBA-1 and the {\em control} can be accounted for by the startup phase, when are just starting to watch a new video (or if we seek to a new point).  The playback buffer starts out empty and carries no useful information to help us choose a video rate. BBA-1 (and BBA-0)  follow the usual rate-map,  starting out slowly (because the buffer is empty) gradually increasing the rate as the buffer fills, as shown in the red line in Figure~\ref{fig:startup_timeseries}. BBA-1 is too conservative during startup. The network can sustain a much higher video rate, we just do not know about it yet. 

Commercial ABR algorithms (including {\em control}) take a gamble and ramp up the video rate much faster than BBA-1, taking the risk that there is more capacity available than they know about. This is a risky game and increases the chances of rebuffering. Indeed, we have observed that for commercial services disproportionate fraction of rebuffer events take place during startup. 

We would like to make our buffer-based algorithms more aggressive during startup (to increase the video rate for those with faster networks), yet carefully understand and manage the risk we are taking on. Our approach is to once again use the buffer occupancy. Specifically, we evaluate the risk of increasing the rate by looking at how much the buffer increases during each chunk download. If a four second chunk downloads in 500ms, we are not taking much risk by doubling the video rate. But we have to be careful. If we increased the video rate by a factor of eight, we would be sailing close to the wind and risk rebuffering. 

Our next algorithm, BBA-2, tries to be more aggressive during the startup phase. It is somewhat analogous to the {\em slow-start} algorithm in TCP; we ramp up quickly to try and find the steady-state rate. At time $t=0$, the buffer is empty. If the first chunk (size $V$ seconds) downloads in less than $0.125 V$s then it downloaded more than eight times faster than it was played, and the buffer increases by more than $0.875 V$s. In this case, BBA-2 picks the next highest video rate, whose chunk size can be {\em four} times bigger. This is a conservative choice, reflecting the lack of knowledge in the system - there is only one chunk in the buffer. As the buffer starts to fill, we have more confidence that it is safe to ramp up the video rate, and we can be more aggressive. As the buffer grows and we have more information, we let BBA-2 increase the video rate faster. Whereas to start with it only increases the video rate if the chunk downloads {\em eight times} faster than it is played, by the time it fills the cushion, BBA-2 is prepared to step up the video rate if the chunk downloads {\em twice} as fast as it is played. The threshold decreases linearly from the first chunk until the cushion is full. 

The blue line in Figure~\ref{fig:startup_timeseries} shows BBA-2 ramping up faster. BBA-2 continues to use the startup algorithm until: (1) the buffer is decreasing, or (2) the buffer occupancy enters the safe zone (i.e., the chunk map suggests a higher rate). 
Afterwards, we use the chunk-map-based algorithm.



\begin{figure}[t]
\centering
\includegraphics[width=\linewidth]{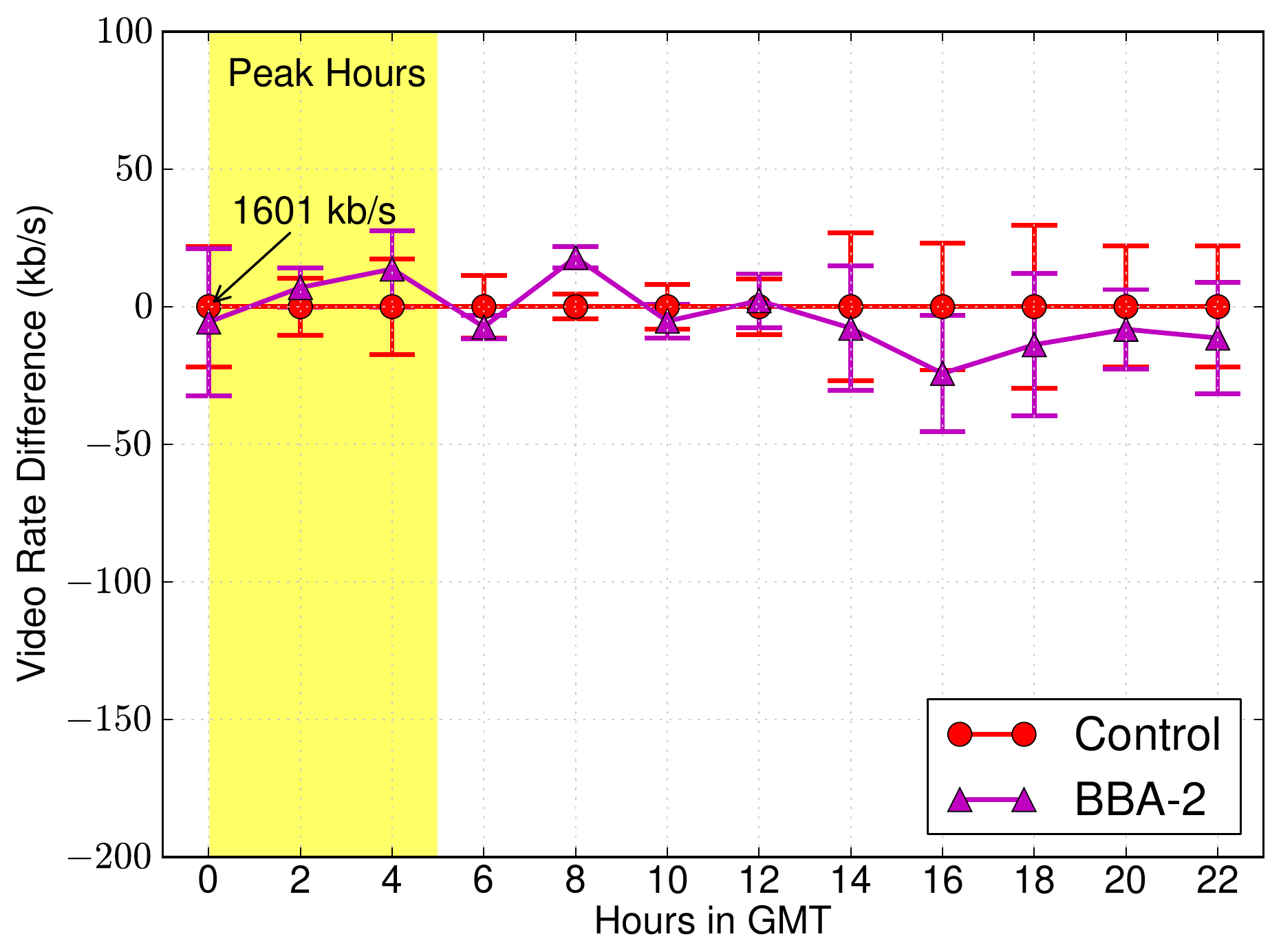}
\caption{BBA-2 achieved similar video rate as the control algorithm.}
\label{fig:sop_vrate}
\end{figure}


\begin{figure}[t]
\centering
\includegraphics[width=\linewidth]{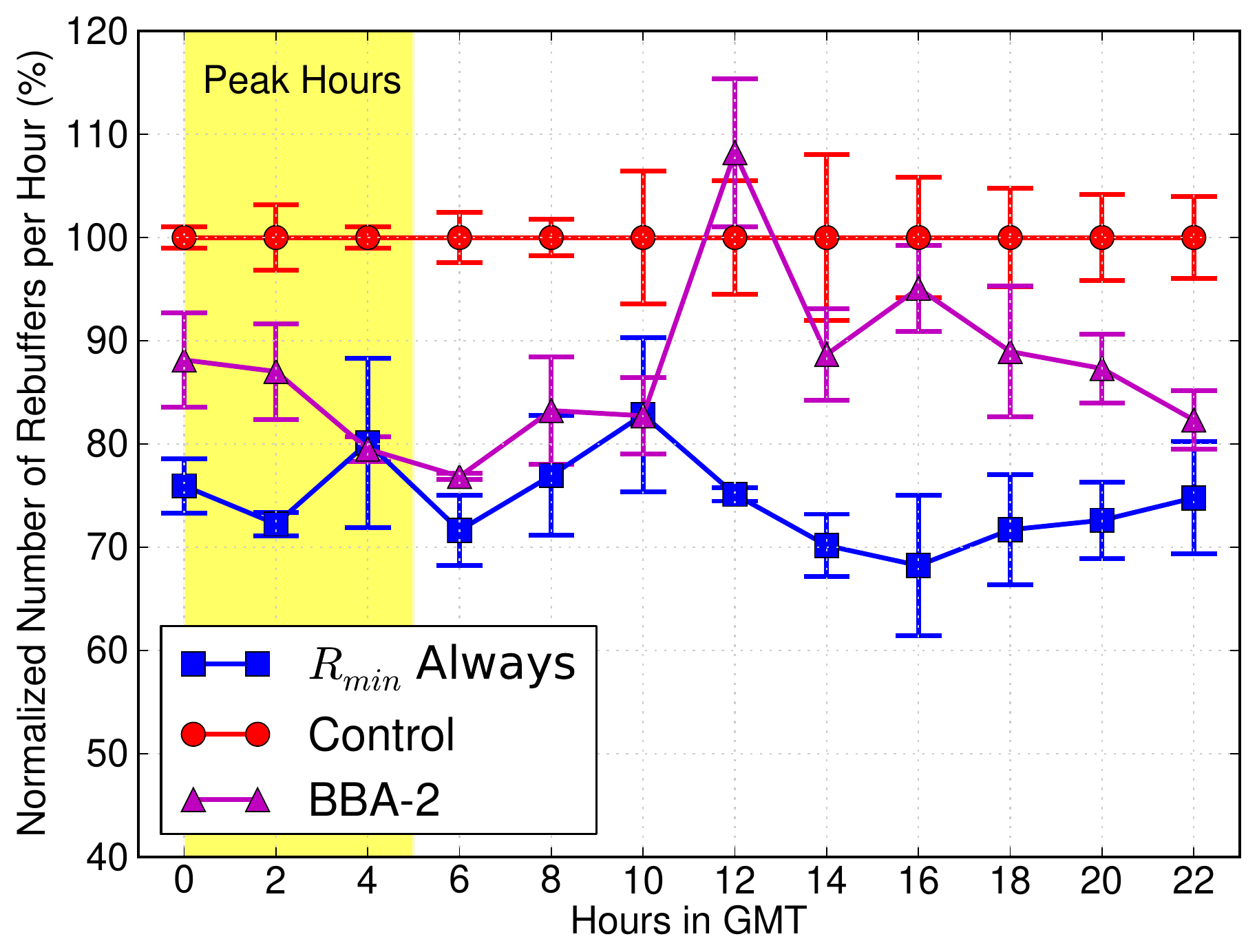}
\caption{BBA-2 has a slightly higher rebuffer rate compare to BBA-1, but still achieved 10-20\% improvement compare to the control during peak hours. Note the numbers are normalized to the average rebuffer rate of the control group for each window.}
\label{fig:sop_rebuf}
\end{figure}



\subsection{Results}
We ran experiments with the same number of users using BBA-2 over the weekend of September 6th to 9th, 2013 along with our experiments with BBA-0 and BBA-1.

Figure~\ref{fig:sop_vrate} shows that BBA-2 does indeed increase the video rate. After optimizing during the startup phase, the video rate with BBA-2 is almost indistinguishable from the {\em control} algorithm. This supports our observations that the lower video rate seen by BBA-0 and BBA-1 were almost entirely due to their conservative rate selection during startup.

Figure~\ref{fig:sop_rebuf} shows that BBA-2 slightly increases the rebuffer rate.  BBA-2 operates in the ``risky'' zone of Figure~\ref{fig:ratemap_concept} and therefore will inevitably rebuffer more often than BBA-1 which only operates in the ``safe'' zone. The good news is that BBA-2 maintains a 10 - 20\% improvement in rebuffer rate compared to the control algorithm during peak hours.


\section{Other Practical Concerns}
\label{sec:others}

\subsection{Handling Temporary Network Outage}
\label{sec:outage}

We have shown that buffer-based algorithms never need to rebuffer if the network capacity is always higher than $R_{\min}$. In this section we explore what happens if network capacity falls {\em below} $R_{\min}$, for example during a complete network outage. Temporary network outages of 20-30s are common, for example when a DSL modem retrains or a WiFi network suffers interference.  To make buffer-based algorithms resilient to brief network outages, we can reserve part of the buffer by shifting the rate-map curve further to the right. 

\begin{figure}[t]
\centering
\includegraphics[width=\linewidth]{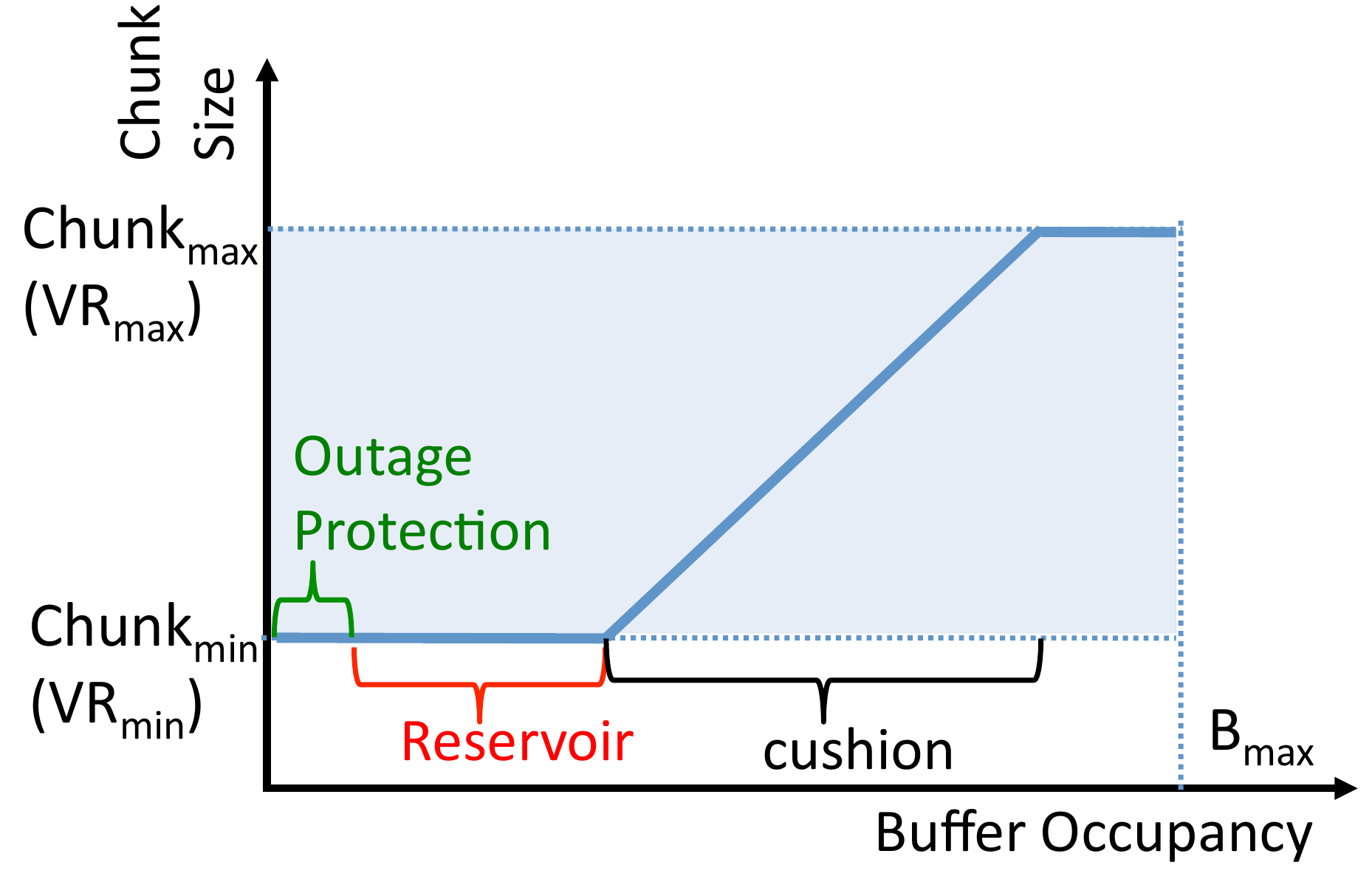}
\caption{To protect against temporary network outage, we allocate part of the buffer as {\em outage protection}.}
\label{fig:chunkmap_realworld}
\end{figure}

Figure~\ref{fig:chunkmap_realworld} shows the chunk map with outage protection. The buffer will now converge to a higher occupancy than before, providing some protection against temporary network outage. 
We call this extra portion of buffer the {\em outage protection}. 

How should we allocate buffers to {\em outage protection}? One way is to gradually increase the size of {\em outage protection} after each chunk is downloaded. 
In the implementation of BBA-1, we accumulate {\em outage protection} by $400$ms for each chunk downloaded when the buffer is increasing and still less than $75$\% full. 
In the implementation of BBA-2, we only accumulate {\em outage protection} after the algorithm exits the startup phase and is using the chunk-map algorithm. 
A typical {\em outage protection} is $20$ - $40$ seconds at steady state and is bounded at $80$ seconds. 
The downside of this approach is that the chunk map keeps moving and can cause video rates to oscillate.
 
In the following, we describe an alternative way to protect against temporary network outage, while reducing changes to the chunk map, by combining it with the dynamic reservoir calculation.


\subsection{Smoothing Video Switch Rate}
\label{sec:changerate}


\begin{figure}[t]
\centering
\includegraphics[width=\linewidth]{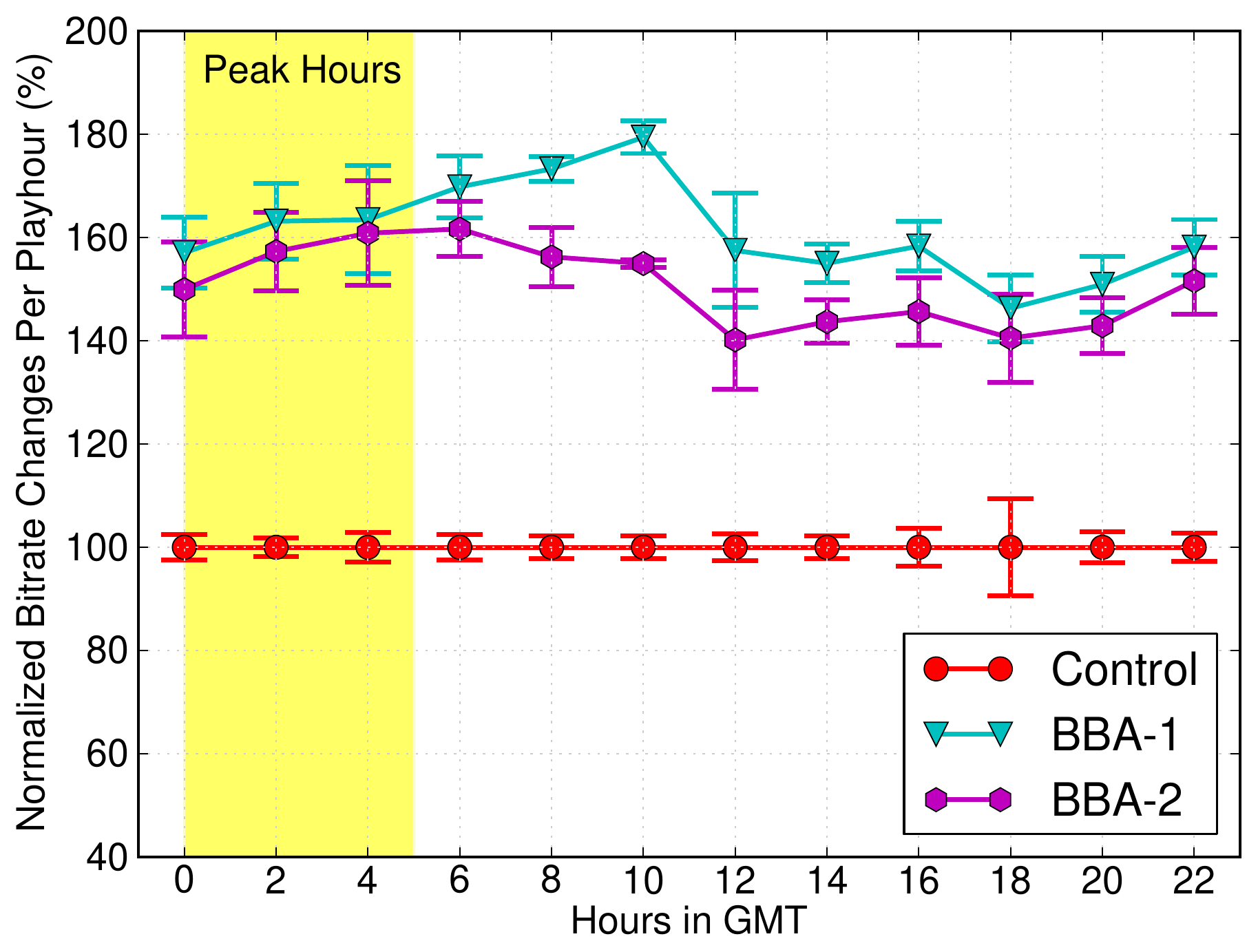}
\caption{After switching from using rate map to chunk map, the video switching rate of BBA-1 and BBA-2 is much higher than the control.}
\label{fig:sop_crate}
\end{figure}

\begin{figure}[t]
\centering
\includegraphics[width=\linewidth]{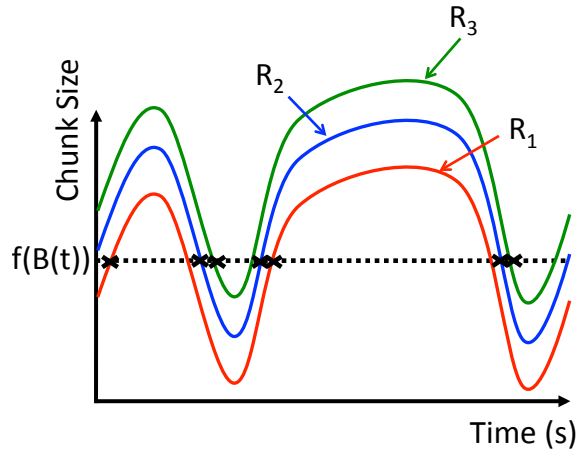}
\caption{When using chunk-map, even if the buffer level and the mapping function remains constant, the variation on chunk sizes in VBR streams can make a buffer-based algorithm switch between rates. The lines on the figure represents the chunk size over time from three video rates, $R_1$, $R_2$, and $R_3$. The crosses represent the points where the mapping function will suggest to change a rate. }
\label{fig:vbr_n_crate}
\end{figure}

By using a chunk map and dynamic reservoir calculation, we showed in Section~\ref{sec:vbr} that we can improve the video rate.
However, this makes the video rate change frequently, as shown in Figure~\ref{fig:sop_crate}.
Although it is debatable whether change rate really matters to the viewer's quality of experience (e.g. if a service offers closely-spaced video rates, the viewer might not notice a switch) we explore mechanisms to reduce the change rate. We will see that by smoothing the changes we can at least match the switching rate of the {\em control} algorithm. 

There are two main reasons why our buffer-based algorithms increase how often we switch to a new video rate. 
First of all, when we use the chunk map, there is no longer a fixed mapping function between buffer levels and video rates. Instead, buffer levels are mapped to chunk sizes and the nominal rate might change every time we request a new chunk.  Even if the buffer level remains constant, the chunk map will cause BBA-1 to frequently switch rates, since the chunk size in VBR encoding varies over time, as illustrated in Figure~\ref{fig:vbr_n_crate}. We reduce the chance of switching to a new rate - and then switching quickly back again - by looking ahead to future chunks. When encountering a small chunk followed by some big chunks, even if the chunk map tells us to step up a rate, our new algorithm {\em BBA-Others} will not do so to avoid possible stepping down.  
The further it looks ahead, the more we smooth out the changes and the less likely the rate will change. If, in the extreme, we look ahead to the end of the movie, it is the same as using a rate-map instead of a chunk-map. Note that BBA-Others only smooths out {\em increases} in video rate. It does not smooth decreases so as to avoid increasing the likelihood of rebuffering. 

Second, the size of the reservoir is calculated from the chunk size variation in the next 480 seconds, as illustrated in Figure~\ref{fig:reservoir_calc}. As a result, the reservoir will shrink and expand depending on the size of upcoming chunks: If large chunks are coming up the chunk map will be right-shifted, and if small chunks are coming up the chunk map will be left-shifted. Even if the buffer level remains constant, a shifted chunk map might cause the algorithm to pick a new video rate. On top of this, as we mentioned in Section~\ref{sec:outage}, a gradual increase in outage protection will {\em also} gradually right-shift the chunk map.  And so we reduce the number of changes by only allowing the chunk map to shift to the right, never to the left; i.e. the reservoir expands but never shrinks. This means the reservoir grows faster than it needs to, letting us use the excess for outage protection. 

\begin{figure}[t]
\centering
\includegraphics[width=\linewidth]{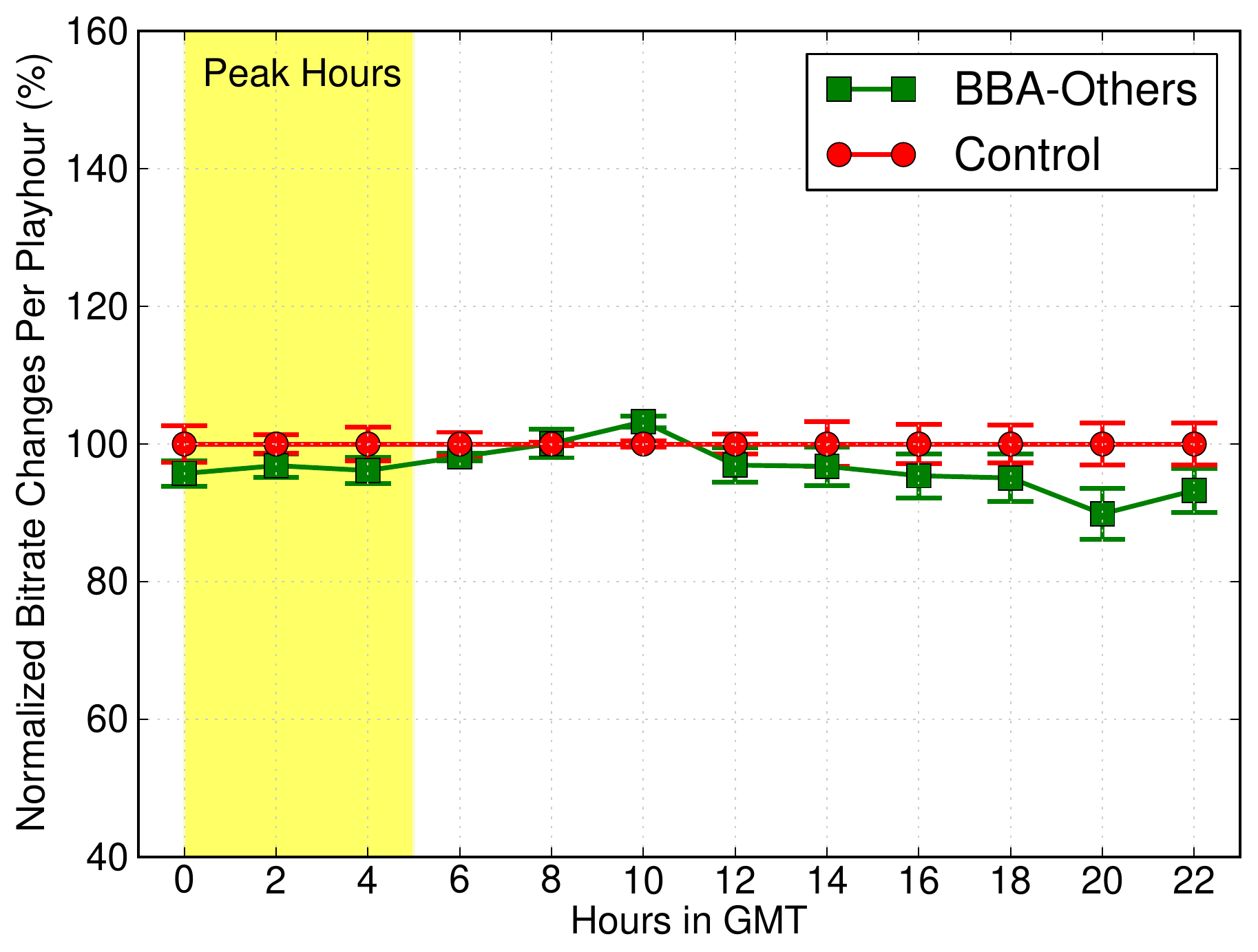}
\caption{BBA-Others smooths the frequency of changes to the video rate, making it similar to the {\em control} algorithm. The numbers are normalized to the average change rate of the control group in each window.}
\label{fig:crate_la}
\end{figure}

\begin{figure}[t]
\centering
\includegraphics[width=\linewidth]{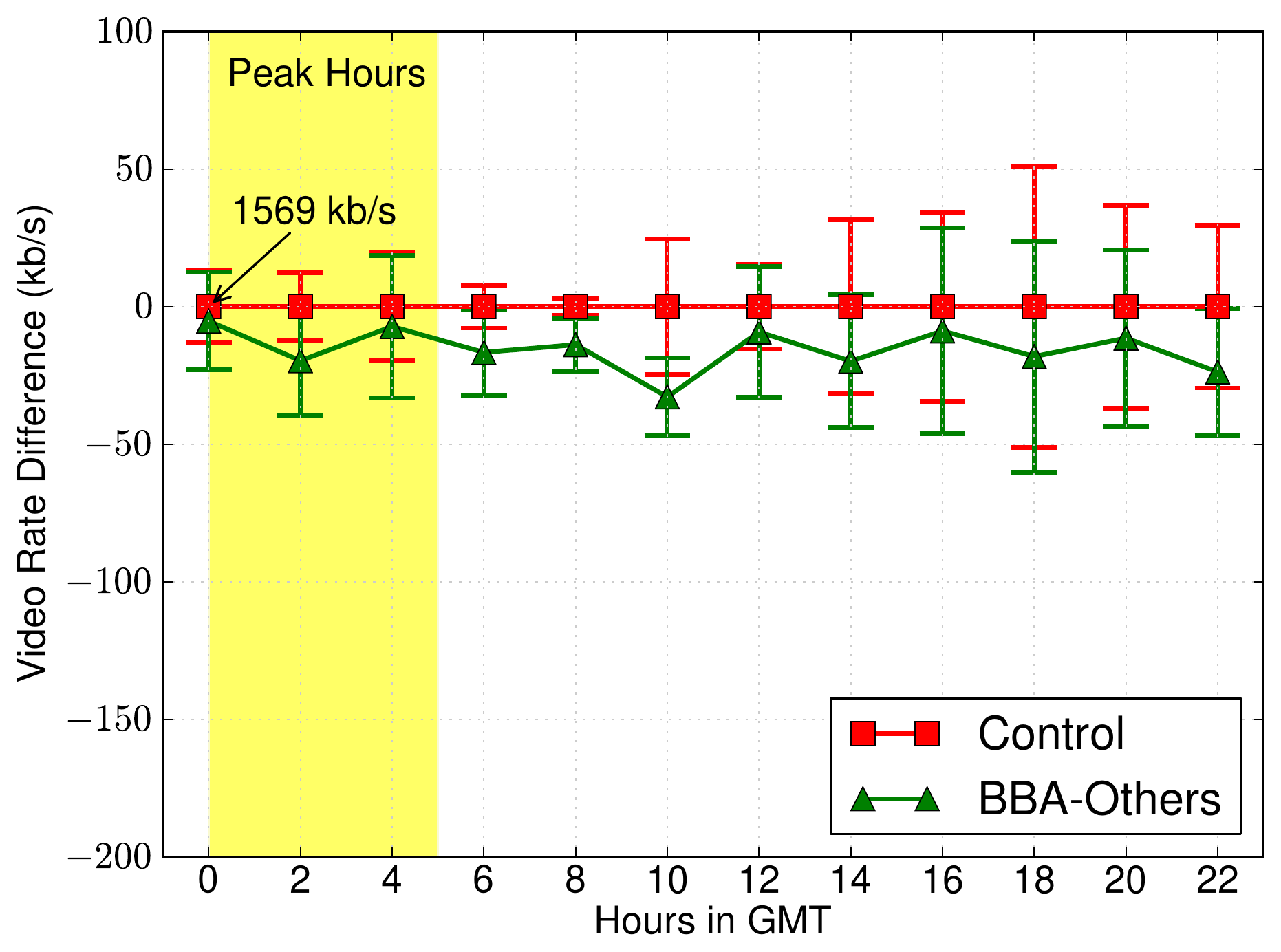}
\caption{BBA-Others achieves similar video rate during the peak hours, but reduces the video rate by 20 - 30 kb/s during the off-peak.}
\label{fig:vrate_la}
\end{figure}

\subsection{Results}
As before, we randomly pick three groups of Netflix users for our experiment.  One third are in the control group, one third always stream at $R_{\min}$, giving us a lower bound on rebuffer rate, and one third  run the BBA-Others algorithm to smooth the switching rate. The experiment was conducted over the weekend of September 20th to 22nd, 2013. 

Figure~\ref{fig:crate_la} shows that the video rate changes much less often with BBA-Others than with BBA-1 or BBA-2 (Figure~\ref{fig:sop_crate}). In fact, BBA-Others is almost indistinguishable from {\em control}; sometimes higher, sometimes lower.\footnote{The numbers are normalized to the average switch rate in the control group for each two-hour window.}
However, although the video rate is almost the same as {\em control}, we  trade about 20kb/s of video rate since we switch up more conservatively. 
Figure~\ref{fig:vrate_la} shows the video rate for BBA-Others. 

\begin{figure}
\centering
\includegraphics[width=\linewidth]{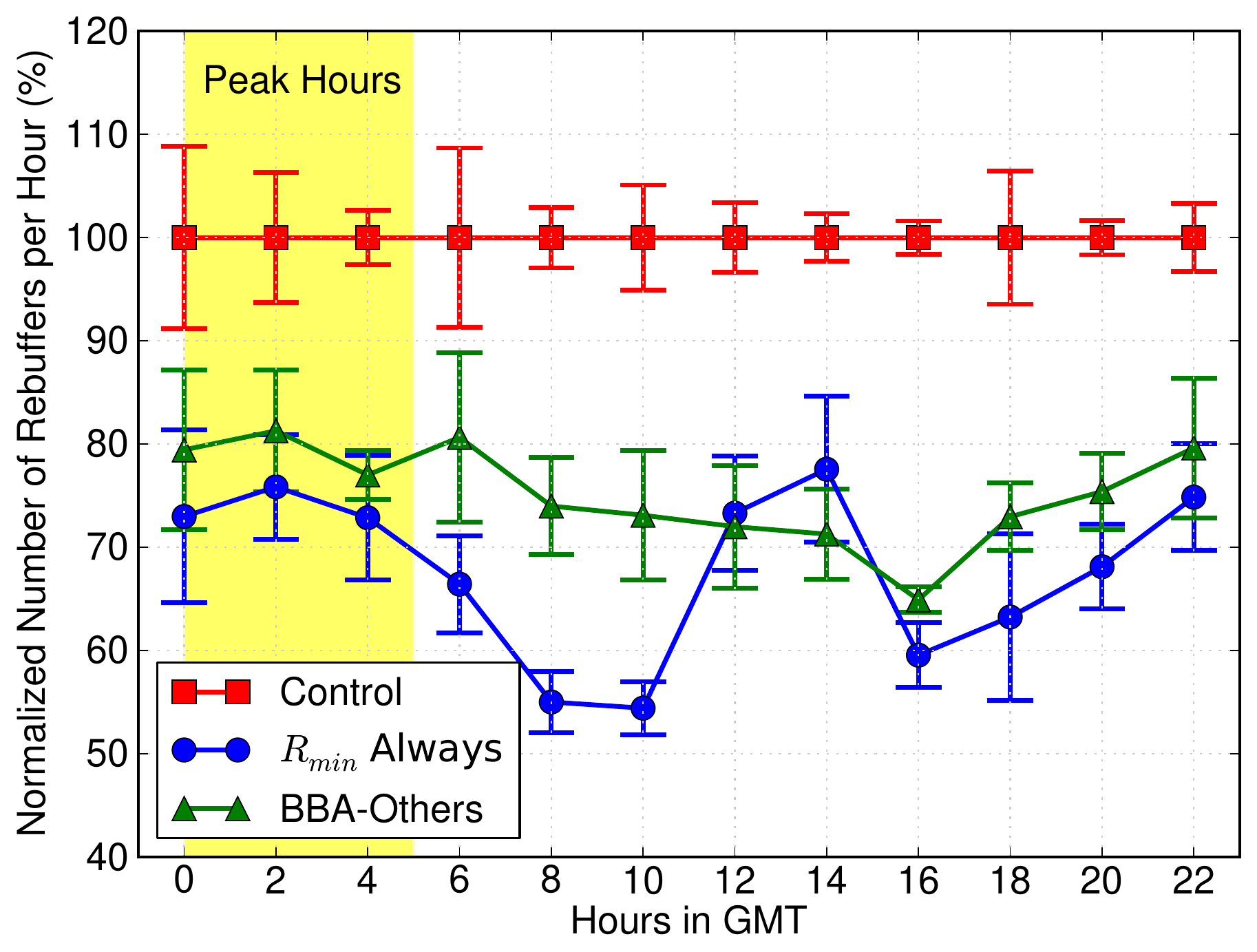}
\caption{{\em BBA-Others} reduces rebuffer rate by 20 - 30\% compare to the {\em control} algorithm. Values are normalized to the average rebuffer rate in the control group for each two-hour window.} 
\label{fig:rebuf_la}
\end{figure}

On the other hand, the rebuffer rate is unaffected --- since we do not change the frequency of switches to a lower rate. BBA-Others maintains the same improvement of rebuffer rate (20 - 30\%) as shown in Figure~\ref{fig:rebuf_la}.


\section{Related Work}
\label{sec:related}

Video streaming services - such as Netflix, YouTube and Hulu - have become very popular in recent years. Because ABR algorithms significantly affect the video quality experienced by users, considerable effort has gone into their design.

{\bf Commercial Streaming Services.} 
While the ABR algorithms differ across commercial video streaming services, the survey in~\cite{markskeynote} describes the general approach followed by the Adobe, Apple and Microsoft http-based services, showing that they all follow similar approaches based on capacity estimation. The same techniques underpin the major commercial services like YouTube, Netflix, and Hulu.
Recently, the MPEG DASH (Dynamic Adaptive Streaming over \http)~\cite{dashspec} standardized the
formatting of video content and leaves open the specific client player ABR algorithm.  

The ABR algorithm used by Netflix is described in~\cite{markskeynote}. Based on recent throughput history, a Netflix client builds a model to predict future bandwidth. It then picks a video rate based on the expected performance of each stream using the prediction model.

{\bf Performance of Commercial Players.}
Several recent measurement papers revealed that accurate bandwidth estimation on top of HTTP is hard~\cite{akhshabi2012happens,huang12imc:videorate,festive_conext12}.
This is because the true network status is hidden several layers below the application layer.
If the video rate is based on an estimate, it can create an undesirable feedback loop and further bias the estimation. 
As a result, players can suffer from poor video quality, unstable rates and unfairness across players.
Jiang et al.~address the unfairness problem by scheduling chunks on top of the rate selection algorithm~\cite{festive_conext12}.
Huang et al.~study approaches to mitigate the estimation bias~\cite{huang12imc:videorate} caused by large chunks.
Ultimately, accurate bandwidth estimation is challenging.  In contrast,
we use the buffer occupancy as a simpler and more robust signal for video rate selection.

\textbf{Rate Selection Algorithms.}
To the best of our knowledge, all previous work on client-side rate selection algorithms is based on bandwidth estimation~\cite{mok2012qdash,xiang2012adaptive,liu2011parallel}. 
The work most similar to ours is~\cite{Tian12conext:buffervideorate}, where the authors design the algorithm as a PID controller with the buffer level as the primary feedback signal. However, their video rate selector still uses a prediction of TCP throughput as an input to the ABR algorithm.

Some work focuses on server-side solutions. YouTube experimented with server-side pacing to ensure full utilization of the link~\cite{trickle,google_keynote_mmsys12}. Among these server-side approaches, the most similar one to ours is \cite{de2011feedback}, where the authors propose an algorithm to maintain the sending buffer at a target level without estimating the bandwidth at all. However, maintaining the sending buffer cannot provide the same rebuffer guarantee as we can at the client side. Ultimately, the client is in the best position to detect and respond to overall dynamics of the system.



\section{Conclusion}
\label{sec:conclusion}
Existing ABR algorithms based their choice on estimates of available network capacity. 
Despite all the complexity these algorithms put into modelling the available capacity, 
our prior work has revealed that estimation-based ABR algorithms can rebuffer unnecessarily.
We conjecture that we can eliminate unnecessary rebuffers by simply {\em picking the video rate as a function of the current occupancy of the playback buffer}.
To validate this, we implemented a variety of buffer-based algorithms in Netflix's browser-based video clients.
We then conduct live experiments with over half a million of Netflix users during May-September 2013.
Our key conclusion is that, at least in the case of clients with large video buffers, 
a novel ABR algorithm design can reduce the rebuffer rate by 20\% compared to an existing best-of-breed ABR algorithm, 
while essentially maintaining a similar video rate.


\balance
\bibliographystyle{abbrv}
\bibliography{paper}

\end{document}